\documentclass[superscriptaddress,prl,twocolumn,tightenlines,10pt,a4paper]{revtex4}%

\usepackage{bbm}
\usepackage{amsfonts}
\usepackage{amssymb}
\usepackage{graphicx}
\usepackage{subfigure}
\usepackage{savesym}
\usepackage{amsmath}
\usepackage{txfonts}
\usepackage{multirow}
\usepackage{epstopdf}
\usepackage{color}

\newcommand{\ket}[1]{\vert#1\rangle}
\newcommand{\bra}[1]{\langle#1\vert}
\usepackage{pbox}
\usepackage{amsmath}
\DeclareMathOperator{\Tr}{Tr}

\begin{document}

\title{Measurement-based quantum computation with trapped ions}

\author{B. P. Lanyon}
\author{P. Jurcevic}

	\affiliation{Institut f\"ur Quantenoptik und Quanteninformation,\\
	\"Osterreichische Akademie der Wissenschaften, Technikerstr. 21A, 6020 Innsbruck,
	Austria}
	\affiliation{
	Institut f\"ur Experimentalphysik, Universit\"at Innsbruck,
	Technikerstr. 25, 6020 Innsbruck, Austria}

\author{M. Zwerger}

	\affiliation{
	Institut f\"ur Theoretische Physik, Universit\"at Innsbruck, Technikerstr. 25, 6020 Innsbruck, Austria
	}

\author{C. Hempel}

	\affiliation{Institut f\"ur Quantenoptik und Quanteninformation,\\
	\"Osterreichische Akademie der Wissenschaften, Technikerstr. 21A, 6020 Innsbruck,
	Austria}

	\affiliation{
	Institut f\"ur Experimentalphysik, Universit\"at Innsbruck,
	Technikerstr. 25, 6020 Innsbruck, Austria}

\author{\\E. A. Martinez}

	\affiliation{Institut f\"ur Quantenoptik und Quanteninformation,\\
	\"Osterreichische Akademie der Wissenschaften, Technikerstr. 21A, 6020 Innsbruck,
	Austria}

	\affiliation{
	Institut f\"ur Experimentalphysik, Universit\"at Innsbruck,
	Technikerstr. 25, 6020 Innsbruck, Austria}

\author{W. D\"ur}

	\affiliation{
	Institut f\"ur Theoretische Physik, Universit\"at Innsbruck, Technikerstr. 25, 6020 Innsbruck, Austria
	}

\author{H. J. Briegel}

\affiliation{Institut f\"ur Quantenoptik und Quanteninformation,\\
	\"Osterreichische Akademie der Wissenschaften, Technikerstr. 21A, 6020 Innsbruck,
	Austria}
	
	\affiliation{
	Institut f\"ur Theoretische Physik, Universit\"at Innsbruck, Technikerstr. 25, 6020 Innsbruck, Austria
	}

\author{R. Blatt}
\author{C. F. Roos}
	
	\affiliation{Institut f\"ur Quantenoptik und Quanteninformation,\\
	\"Osterreichische Akademie der Wissenschaften, Technikerstr. 21A, 6020 Innsbruck,
	Austria}
	
	\affiliation{
	Institut f\"ur Experimentalphysik, Universit\"at Innsbruck,
	Technikerstr. 25, 6020 Innsbruck, Austria}

\date{\today}

\begin{abstract}

Measurement-based quantum computation (MBQC) represents a powerful and flexible framework for
quantum information processing, based on the notion of entangled quantum states as computational resources.
The most prominent application is the one-way quantum computer, with the cluster state as its universal resource.
Here we demonstrate the principles of MBQC using deterministically generated graph states of up to 7 qubits, in a system of trapped atomic ions.
Firstly we implement a universal set of operations for quantum computing. Secondly we demonstrate a family of measurement-based quantum error correction codes, and show their improved performance as the code length is increased.
We show that all our graph states violate a multipartite Bell inequality and are therefore capable of information processing tasks that cannot be described by a local hidden variable model. The methods presented can directly be scaled up to generate graph states of several tens of qubits.

\end{abstract}

\maketitle

The circuit model of quantum computation is conceptually similar to a classical computer: a register of two-level systems in a simple initial product state are manipulated using unitary quantum logic gates \cite{MikeIke}. MBQC \cite{PhysRevLett.86.5188} represents a conceptually and practically different approach: after preparing an entangled cluster state of qubits \cite{PhysRevLett.86.910},   computation proceeds by performing measurements and feedforward. Both approaches present different theoretical and practical challenges to realisation and warrant investigation in parallel.

Recently, researchers have found novel applications for MBQC beyond universal QC, including e.g. blind quantum computation \cite{blind, Barz20012012} measurement-based entanglement purification \cite{PhysRevLett.110.260503} and quantum error correction \cite{Zwerger2013new}, featuring very high thresholds. Owing to the two-stage process of MBQC -- resource creation followed by its processing -- resources states can be purified and manipulated beforehand. This offers a large degree of flexibility in optimizing and compressing schemes for quantum information processing.
Schemes for correcting errors in universal MBQC have also been found with extremely high tolerence to errors, compared to those known for the circuit model of QC \cite{PhysRevLett.98.190504, Knill:2005ly, Zwerger2013new}.

Important experimental progress on MBQC has been made using entangled states of up to 8 photonic qubits \cite{Walther:2005fk, Prevedel:2007uq, Yao:2012fk}. Scaling up the non-deterministic methods used to generate entangled states in these works is very challenging, since their success probably reduces exponentially in photon number. Very recently there has been work on generating cluster states in continuous variables of light fields \cite{continuousclusters}.

In this work we present the first demonstration of MBQC using trapped ions. Furthermore, we make two experimental steps forward in the model of MBQC that are system-independent: the deterministic generation of cluster states and the demonstration of quantum error correction (QEC). 
The paper is organised as follows; firstly MBQC is briefly reviewed and our approach to preparing cluster states is summarised; then a universal set of operations is presented using a 4 qubit cluster state; finally the phase-flip correction code is demonstrated for increasing codeword lengths. We do not implement active feedforward, which has previously been demonstrated with trapped ions \cite{Riebe:2004, Riebe:2008uq}. Our results are post-processed to reproduce the action of perfect feedforward. 

A mathematical graph $G=(V,E)$ is a set of vertices $V$ and edges $E$.
The corresponding \emph{graph state} is a physical state of $n=|V|$ qubits, associated with the vertices of the graph $G$, which is defined in the following way.
For every vertex one defines an operator $K_a=X_a\prod_{b\in {\cal N}(a)} Z_b$ where ${\cal N}(a)$ denotes the neighborhood of vertex $a$ and $X$ and $Z$ denote Pauli spin-$\frac{1}{2}$ operators.
The graph state $| G \rangle$ is then uniquely defined as the common eigenstate of all operators $K_a$ with eigenvalue $+1$.
One approach to generating graph states is to start from an initial state with all qubits in $\ket{+}$, i.e. the +1 eigenstate of $X$, and then apply a controlled phase (CP) gate \cite{MikeIke} between every pair of qubits connected by an edge.

An important graph state is the 2D cluster state $\ket{C}$ \cite{PhysRevLett.86.910}, which has the topology of a square lattice and belongs to the class of universal resource states \cite{PhysRevLett.97.150504, NJO_briegel}. Any quantum computation can be carried out on a sufficiently large $\ket{C}$. In particular, any quantum logic circuit can be translated to a single-qubit-measurement pattern on $\ket{C}$   \cite{PhysRevLett.86.910}. Measurements in the computational basis can be used to remove qubits from the cluster and imprint any desired quantum circuit structure onto the lattice. Measurements in the basis $B(\alpha)=\left(\ket{+\alpha},\ket{-\alpha} \right)$, where $\ket{\pm \alpha}=\left(\ket{0}\pm e^{i\alpha}\ket{1}\right)/\sqrt{2}$ with real $\alpha$, drive the computation. The value of $\alpha$ determines, for example, the angle of single-qubit rotations. In general $\alpha$ has to be adapted to the outcomes of previous measurements and thus introduces a temporal order and the need for feedforward \cite{PhysRevLett.86.5188}.

While the use of the state $\ket{C}$ offers a canonical route to quantum circuit simulation, the conceptual framework of MBQC is much more flexible and offers different and more economic modes of operation. Instead of starting with $\ket{C}$, for example, one can use graph states as resources which are tailored for specific quantum circuits, which could also be subroutines of a larger quantum algorithm. For circuits that contain mainly Clifford gates (including e.g.\ the CNOT and the Hadamard gate) their resource states can be highly compressed without loss of functionality. Such a situation arises for example in QEC, which will be described below. Concatenating such tailored resource states in measurement-based QEC and entanglement purification can give rise to much higher thresholds \cite{PhysRevLett.110.260503, Zwerger2013new}

Experiments use strings of $^{40}$Ca$^+$ ions in a linear Paul trap, manipulated with lasers. Two electronic states encode a qubit ($|D_{5/2},m=+3/2\rangle{=}\ket{0}$, $|S_{1/2},m=+1/2\rangle{=}\ket{1}$) that are coupled by an electric quadrupole transition at 729~nm. 
We now briefly summarise how graph states are generated, for more details see the supplementary material. 
Experiments begin by preparing $n$ ionic-qubits in the product state $\ket{1}^{\otimes n}$ via optical pumping, and preparing the axial centre-of-mass (COM) and stretch (STR) vibrational modes in the ground state by resolved sideband cooling. Graph states are generated using three distinct tools. Firstly an effective long-range qubit-qubit interaction of the form $H_{\rm MS}=J\sum_{a<b}X_aX_b$ can be turned on for arbitrary times. This interaction is realised by off-resonantly driving the axial COM vibrational mode of the ion string using a bichromatic laser field from a single direction \cite{Kirchmair:2009, PhysRevLett.82.1971}. When applied to ionic-qubits in $\ket{1}^{\otimes n}$ this periodically generates fully connected graph states that are equivalent to GHZ states \cite{PhysRevLett.106.130506, PhysRevLett.82.1835}. The second and third tools both allow the selective removal of any connection in the graph \cite{Riebe:2004,PhysRevA.79.012312}. Both use laser pulses tightly focused on individual ions. In combination these three tools allow, in principle, any cluster state to be deterministically created. 

In the circuit model of QC a universal set of logic gates provides the tools to implement arbitrary quantum algorithms \cite{MikeIke}. A common universal gate set consists of the CP gate and arbitrary single-qubit rotations around two independent axes, e.g. the Z axis, $R_Z(\alpha)=e^{-i\frac{\alpha}{2} Z}$ and  the X axis, $R_X(\alpha)=e^{-i\frac{\alpha}{2} X}$, by angle $\alpha$ \cite{MikeIke}.
All of these gates can be translated to carrying out specific sequences of measurements and feedforward on a four qubit linear cluster state $\ket{LC_4}$, which are presented in figures 1a and 2a \cite{Walther:2005fk}. Realising measurement patterns like these on large-scale cluster states, when combined with QEC, enables arbitrary MBQC  \cite{PhysRevLett.86.5188}. We create $\ket{LC_4}$ using a laser pulse sequence lasting 300$\mu$s  %(to be checked)
 and reconstruct the full density matrix via quantum state tomography (see supplementary material for details).
 The observed fidelity with the ideal state is $0.841\pm{0.006}$,
 which is well above the threshold for multipartite entanglement of 0.5 \cite{PhysRevLett.94.060501}.
%representing a $21\pm{xx}$\% improvement over previous results using photons \cite{Walther:2005fk, Prevedel:2007uq}.
%walter obtained $0.62\pm{0.01)$ 2007
%walter obtained $0.63\pm{0.02)$ 2005

Measurement of $\ket{LC_4}$ in the order presented in figure 1 is equivalent to a circuit performing a sequence of one-qubit gates on the encoded state $\ket{+}$. The choice of measurement basis of qubits 1,2 and 3 [$B_1(\alpha)$, $B_2(\beta)$ and $B_3(\gamma)$] determines the overall rotation applied to $\ket{+}$. We implement a range of measurement combinations, each demonstrating a different one-qubit rotation. One approach, which avoids the need for active feedforward, is to reconstruct the output state (encoded in qubit $4$) post-selected on the cases where the $+1$ outcomes of the measurement of qubits 1, 2 and 3 are observed, as in \cite{Walther:2005fk}.   More information is obtained if all outcomes are kept and post-processed to simulate perfect feedforward. Results obtained in this way provide an upper limit for the performance that could have been achieved using feedforward. 
Figure 1 presents the results on the Bloch sphere: a range of different rotated output states, reconstructed via quantum state tomography. The average output state fidelity with the ideal case is $0.92\pm{0.01}$.

Measurement of $\ket{LC_4}$ in the order presented in figure 2 is equivalent to a circuit composed of a CP gate and one-qubit rotations, which operates on the encoded state $\ket{++}$. The choice of measurement basis' of qubits 1 and 4 determine the one-qubit rotations. The output state is stored in qubits 2 and 3. We choose two important cases: a maximally entangled state (case 1) and a product state (case 2) are ideally created. We quantify the generated entanglement by the tangle $\tau$ \cite{PhysRevA.61.052306}.
In case 1 we find that the experimentally reconstructed state is strongly entangled, $\tau{=}0.59\pm{0.05}$, and has a fidelity of $0.88\pm{0.02}$ with the ideal state. In case 2 the experimental state is close to being separable,  $\tau{=}0.02\pm{0.01}$, and has a fidelity of $0.83\pm{0.01}$ with the ideal state. Experimentally reconstructed two-qubit output density matrices are presented in figure 2. Taken together, the results in figures 1 and 2 demonstrate a universal set of operations.

We also created the 4-qubit box cluster state  $RC_4$, an example of a ring cluster \cite{PhysRevLett.95.120405} and the smallest intrinsically 2D cluster state. We tomographically reconstructed the full density matrix and observed a fidelity of $0.847\pm{0.007}$ with the ideal state. For more details see supplementary material.

In a realistic setup one cannot decouple the qubits on which the computation is performed completely from the environment, which will introduce errors on the qubits. QEC codes \cite{PhysRevA.54.1098, Steane08111996, MikeIke} provide a solution by encoding the states $\ket{0}$ and $\ket{1}$ of a qubit into the states of larger physical systems $\ket{0_L}$ and $\ket{1_L}$ (called codewords or logical qubits) and using correlations to protect the information. QEC consists of at least two steps after the encoding. Firstly, one measures the correlation operators which reveal the error syndrome. In this step errors, which might be unitary qubit rotations or involve entangling to the environment, are discretized. The discretization of the errors is a crucial step, as it reduces the infinite set of possible quantum errors to a finite set. Secondly, one applies the recovery operator to undo the error.  The principles of QEC in the circuit model have been demonstrated before \cite{PhysRevLett.81.2152, Chiaverini:2004vn, PhysRevA.71.052332, superQEC, PhysRevLett.109.100503}, including the 3-qubit phase-flip code \cite{Schindler27052011}. Circuit model QEC codes can be translated to MBQC in the same way as algorithms. However, it is important to note that QEC codes involve only Clifford gates and Pauli measurements, which can be implemented in a very compact way in MBQC \cite{Zwerger2013new}.

We demonstrate an MBQC phase-flip code, with codewords $\ket{0_L}=\ket{+}^{\otimes n}$ and $\ket{1_L}=\ket{-}^{\otimes n}$, capable of correcting full phase flips (Z) on up to $(n-1)/2$  of the codeword qubits. 
The general form of the graph state employed, labelled $\ket{EC_n}$, and the protocol are presented in figure 3. $\ket{EC_n}$  consists of $n{+}2$ qubits: $n$ for the codeword, labelled $C_1$ to $C_n$, and two additional qubits $A$ and $B$ to read in (encode) and read out the initial and final protected 1-qubit state, respectively. For arbitrary $n$,
\begin{equation}
2\ket{EC_n}{=}(\ket{0}_A\ket{0_L}+\ket{1}_A\ket{1_L})\ket{0}_B+(\ket{0}_A\ket{1_L}+\ket{1}_A\ket{0_L})\ket{1}_B
\end{equation}
After preparation of $\ket{EC_n}$ , the protocol proceeds as follows:
1. A 1-qubit state $\ket{\psi}$ or the orthogonal state is encoded by measuring qubit A in a basis where these are the eigenstates.  The effect is to distribute either state non-locally amongst the remaining $(n+1)$-qubits.
2. Each of the $n$ central qubits $C_n$ is measured in the X basis, yielding one of $2^n$ possible outcomes. This simultaneously decodes the state and reveals which of up to $(n-1)/2$  errors have occurred on the central qubits (i.e. determines the error syndrome).
3. A 1-qubit correction operation, determined by the outcome in 2 is applied to the output state, stored in qubit $B$, recovering the encoded 1-qubit state.

The temporal order of the measurements is unimportant and errors can happen to the central qubits, $C_1$ to $C_n$, at any time before measurement of these central qubits. It is useful to interpret the protocol as attempting to teleport a state across the cluster, from A to B, through a noisy channel affecting the middle qubits. 

We demonstrate the protocol using the $n=$1, 3 and 5 cases shown in figure 4.a-c. Equivalent experimental investigations of increasing codeword lengths in the circuit model have not yet been realised, due to the complexity of the gate sequences required. The laser-pulse sequences used to generate each graph are described in the supplementary material. For the $n=1$ and 3 cases we reconstruct the full $(n+2)$-qubit density matrices via quantum state tomography, yielding state fidelities of $0.92\pm{0.005}$ and $0.843\pm{0.005}$, respectively.

The codes are tested against errors realised by applying 1-qubit rotations $R_z{=}\exp(-i\frac{\theta}{2}Z)$ to all or a subset of the codeword qubits $C_n$. After measurement of $C_n$ (and therefore discretisation of the errors) this is equivalent to independent phase flips occurring incoherently and independently on those qubits to which it is applied, with probability $p=\sin^2(\theta/2)$. For input states we choose to encode the four eigenstates of the Pauli X and Y operators, which are maximally affected by phase flip errors. Error correction performance is quantified by the average teleportation state fidelity (ATF), through the noisy channel, averaged over the four input states.

Firstly each code is tested against errors applied to one codeword qubit ($C_1$). The $n=$ 3 and 5 graphs should be robust to this, whilst the ATF for the $n=$ 1 graph should reduce linearly with $p$, since it provides no error correction.  The results, presented in figure 4, show quantitative agreement with the ideal cases, up to deviations that largely correspond to an overall fidelity drop due to imperfections in the graph state preparation. Also shown is the resistance against two errors ($C_1$ \& $C_2$) for the $n{=}5$ case, afforded by the increased codeword length. We emphasise the quality of the results: even in the presence of large amount of noise we are able to teleport states across a 7 ionic-qubit string with fidelities of over 0.8.

These diagnostic tests show that the experimentally generated graph states respond correctly to errors applied to individual codeword qubits. A more realistic situation is that \emph{all} codeword qubits are subject to error with the same probability. 
Figure 5d shows the theoretical performance of the ideal graphs against such noise: for $p<0.5$, graphs with larger $n$ perform better, tending towards perfect correction up to $p=0.5$ as $n\rightarrow\infty$. This is a challenging experiment since the small improvement offered by the larger ideal graphs is mitigated by the reduced quality with which the larger graphs were generated.

Errors are applied to all physical codeword qubits in the experimentally generated graph states and the results are presented in figure 5e. Qualitative agreement with the ideal case is observed. Even though many more qubits are exposed to errors in the larger codewords, there is still a region where they perform better. That is we are able to demonstrate that, for a range of noise levels, a better protection of quantum information is provided when using a larger error correction code.
For more discussion see supplementary material.

%In order to achieve the reported process fidelities, a high fidelity preparation of the resource states is required. 
Table \ref{stats} summarises important properties of all experimentally generated graph states. 
%We were able to demonstrate the non-local properties of all prepared graph states. 
As shown in \cite{PhysRevLett.95.120405}, any ideal graph state violates a multipartite Bell-type inequality, i.e. it cannot be described by a local hidden variable model. By measuring the expectation value of the multipartite Bell observable ${\cal{B}}_n$, described in the supplementary material, we find a clear violation of the corresponding inequality for all prepared graph states. These results imply that the information processing shown in the experiment is intrinsically quantum, and cannot be described by a local hidden variable model. Furthermore, ${\cal{B}}_n$ is equal to the state fidelity, but requires only a small fraction of the number of measurements for state tomography. Our largest error sources, which limit graph state quality, are laser intensity fluctuations and random electric fields that heat the ion string.

Using ionic-qubits we have made several distinct steps forward in MBQC: the deterministic generation of graph states, together with their application as resources;  the demonstration of measurement-based quantum error correction and; the observation of improved performance with increasing codeword length. 
It should be possible to directly scale-up the presented techniques to generate cluster states of several tens of qubits in the next few years. 
More efficient schemes for generating ionic cluster states might employ direct nearest-neighbour interactions, such as those afforded by multi-site ion traps \cite{Harlander:2011ys, Brown:2011zr, muir, PhysRevA.79.052324}.
Both the circuit and measurement-based models of QC have now been demonstrated in trapped ions. Since all instances have been proof-of-principle experiments, there is as yet no obvious reason to choose one over the other at this stage. Both paths present similar challenges in terms of improvements over our ability to control many-body quantum systems.\\
\\
\noindent BPL acknowledges support by a Marie Curie Fellowship (PIIF-GA-2010-275477).  
This work was supported by the Austrian Science Fund (FWF) under grant numbers: P25354-N20, P24273-N16 and SFB F40-FoQus F4012-N16.

%\bibliography{clusters}

\newpage

\begin{figure*}[t]
\vspace{0mm}
\includegraphics[width=1.3 \columnwidth]{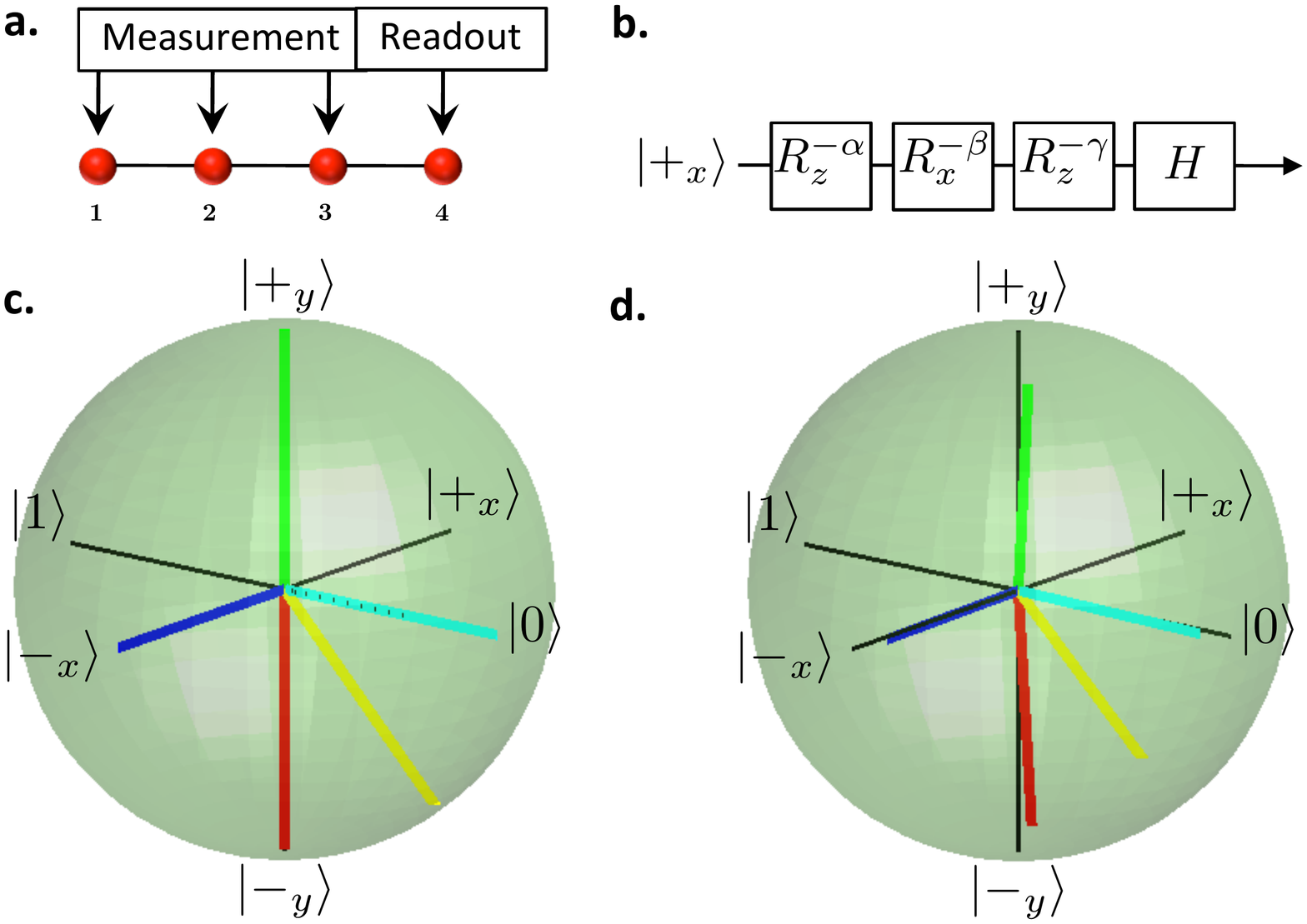}
\vspace{0mm}
\caption{\label{error} \textbf{Demonstration of one-qubit gates via measurement-based quantum computing}. \textbf{a}.
Qubits 1, 2 and 3 of the four qubit linear cluster $\ket{LC_4}$ are measured consecutively in the basis $B_1(\alpha)$, $B_2(\pm{\beta})$ and $B_3(\pm{\gamma})$, respectively, with signs determined by previous measurements outcomes. Qubit 4 encodes the output state.
\textbf{b}. Equivalent quantum circuit, with angles determined by $\alpha$, $\beta$ and $\gamma$. $H=$ Hadamard \cite{MikeIke}. 
\textbf{c}. Ideal output states on the Bloch sphere for $[\alpha,\beta,\gamma]{=}[\pi/2,0,0]$ (red), $[0,0,-\pi/2]$ (green),   $[\pi/2, -\pi/2, 0]$ (blue), $[\pi/2,0,-\pi/2]$ (cyan), $[\pi/4,0,0]$ (yellow).
\textbf{d}. Experimentally measured states, the average fidelity with the ideal cases is $0.92\pm{0.01}$.
}
\label{levelscheme}
\vspace{6mm}
\end{figure*}
%there where actual 9 states produced by several of them are the same.
%the 4 plotted above are stored together in a matlab variable called out_final, produced by:
%/Users/ben/Desktop/clusters/svn/final_DATA/universal_gate_set/one_qubit_gates/figure_plotter/plot_results/
%here is the out_final structure, it uses photon notation to label the vectors:

%out_final =
%
%        order: {'Lm'  'Rm'  'Am'  'Hm'  'Funny'}
%       rhosML: {[2x2 double]  [2x2 double]  [2x2 double]  [2x2 double]  [2x2 double]}
%    rho_ideal: {[2x2 double]  [2x2 double]  [2x2 double]  [2x2 double]  [2x2 double]}
%       rhosmc: {1x5 cell}
%         fids: [0.9450 0.9050 0.8950 0.9550 0.8960]
%     fids_std: [0.0157 0.0195 0.0225 0.0152 0.0199]
%         purs: [1x5 double]
%     purs_std: [0.0295 0.0327 0.0363 0.0285 0.0356]
%       fid_av: 0.9192
%    fid_avstd: 0.0084

\begin{figure*}[t]
\vspace{0mm}
\includegraphics[width=1.2 \columnwidth]{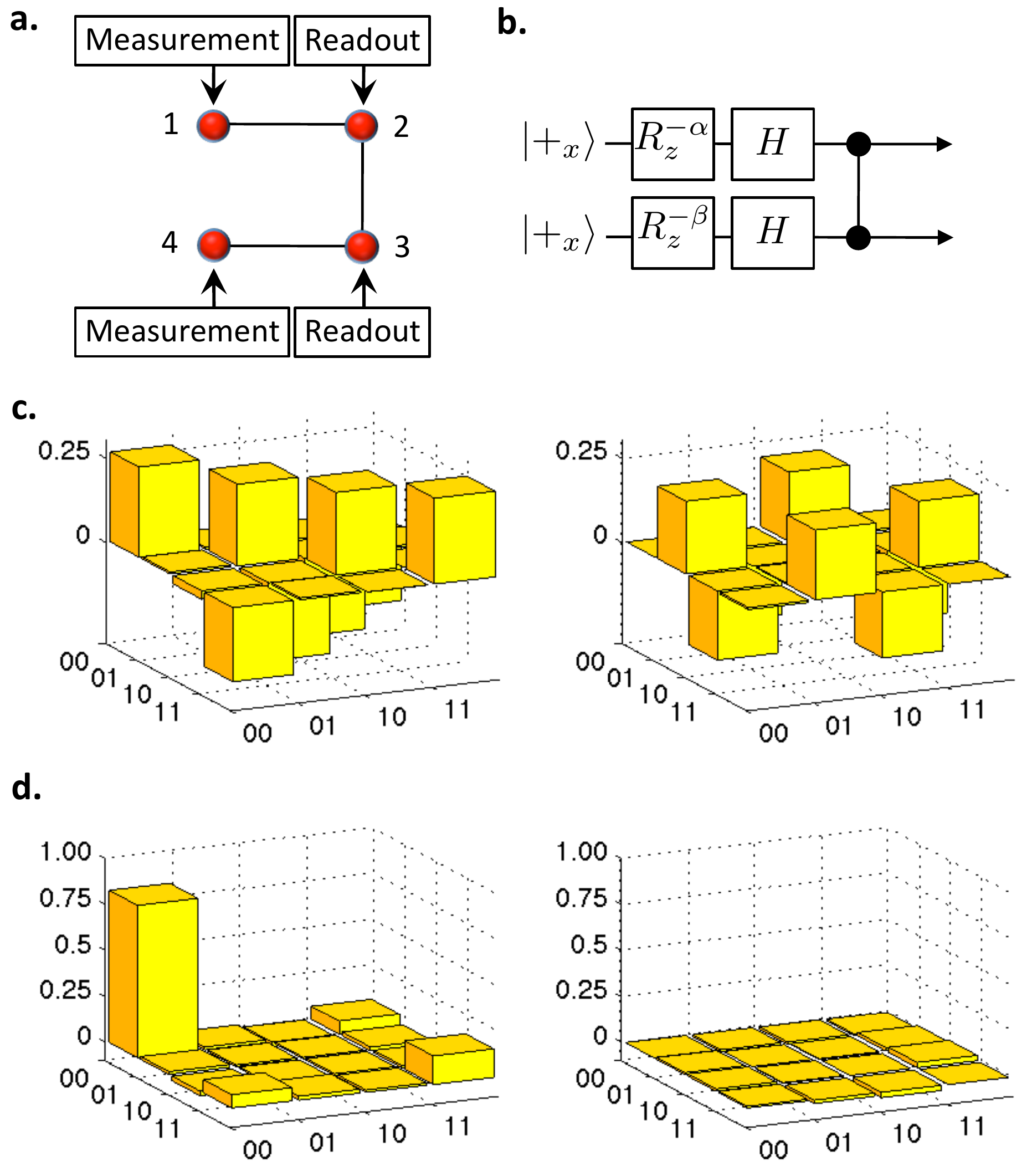}
\vspace{0mm}
\caption{\label{error} \textbf{Demonstration of two-qubit gates via measurement-based quantum computing}. \textbf{a}.
Qubits 1 and 4 of the four qubit linear cluster $\ket{LC_4}$ are measured in the basis $B_1(\alpha)$ and $B_4(\beta)$, respectively. Qubits 2 and 3 encode the output state.
\textbf{b}. Equivalent quantum circuit, with angles determined by $\alpha$ and $\beta$. Two black dots connected with a vertical line is a controlled phase (CP) gate. 
\textbf{c-d}. Experimentally reconstructed output state density matrices (left and right show real and imaginary parts, respectively) in two cases: \textbf{c} An entangled state for $\alpha{=}\pi/2$, $\beta{=}-\pi/2$, with fidelity $0.88\pm{0.02}$, and tangle $0.59\pm{0.05}$;  \textbf{d} An ideally separable state for $\alpha{=}0$, $\beta{=}0$, with fidelity $0.83\pm{0.01}$ and tangle $0.02\pm{0.01}$.
}
\label{levelscheme}
\vspace{4mm}
\end{figure*}
%there where four two qubits state produced 1,2,3,4. We present numbers 1 and 4. Here are the results for all 4, produced by Ben's analysis. see file svn/final_DATA/universal_gate_set/two_qubit_gates/figure_plotter.m

%out =
%
%           rho: {{1x1 cell}  {1x2 cell}  {1x3 cell}  {1x4 cell}}
%          data: {[9x5 double]  [9x5 double]  [9x5 double]  [9x5 double]}
%    projectors: {[4x4x36 double]  [4x4x36 double]  [4x4x36 double]  [4x4x36 double]}
%         rhomc: {[4x4x101 double]  [4x4x101 double]  [4x4x101 double]  [4x4x101 double]}
%        fid_ML: [0.8815 0.8888 0.8761 0.8266]
%        pur_ML: [0.7850 - 0.0000i 0.7984 - 0.0000i 0.7859 - 0.0000i 0.7237 - 0.0000i]
%        tan_ML: [0.5874 0 0.0024 0.0207]
%        fid_MC: {[1x100 double]  [1x100 double]  [1x100 double]  [1x100 double]}
%        pur_MC: {[1x100 double]  [1x100 double]  [1x100 double]  [1x100 double]}
%       tang_MC: {[1x100 double]  [1x100 double]  [1x100 double]  [1x100 double]}
%       fid_std: [0.0167 0.0128 0.0149 0.0097]
%       pur_std: [0.0273 0.0203 0.0244 0.0144]
%      tang_std: [0.0523 0.0024 0.0066 0.0142]
%        ideals: {[4x4 double]  [4x4 double]  [4x4 double]  [4x4 double]}

%so the fids are:
%$0.8815\pm{0.0167}$, $0.8266\pm{0.0097}$
%the tangles are:
%$0.5874\pm{0.0523}$, $0.0207\pm{0.0142}$

\newpage

\begin{figure*}[t]
\vspace{0mm}
\includegraphics[width=1.5 \columnwidth]{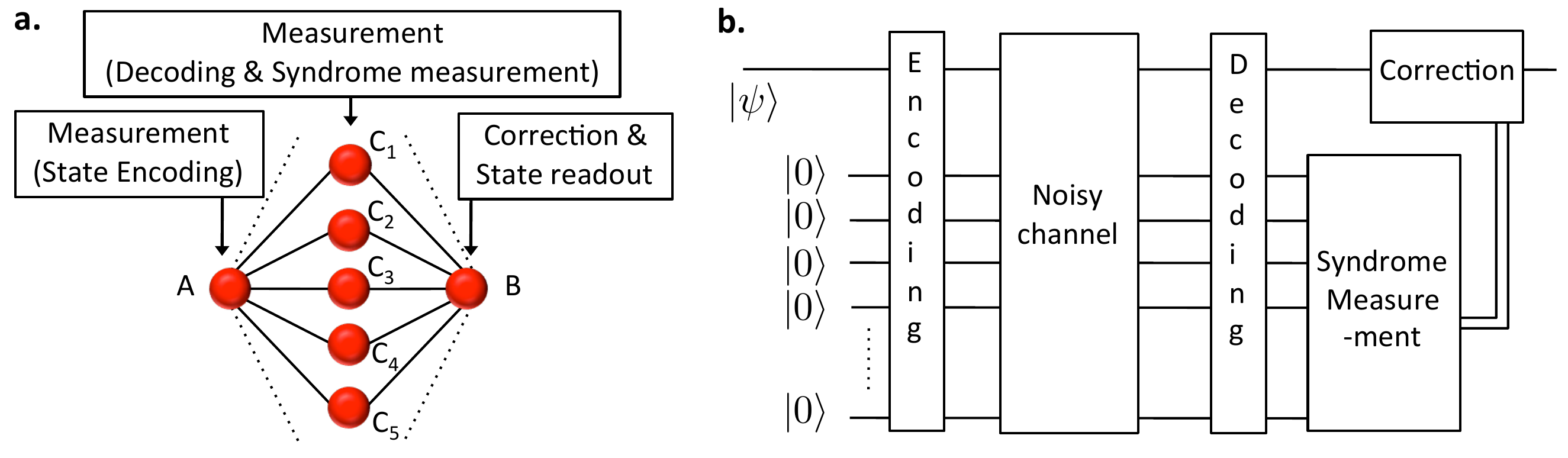}
\vspace{5mm}
\caption{\label{error} \textbf{Graph state $\ket{EC_n}$  and its use in measurement-based quantum error correction}.
\textbf{a}.
$(n+2)$-qubit graph state $\ket{EC_n}$  and protocol, which can correct for phase-flip errors ($Z$) occurring on up to $(n-1)/2$  of the central qubits, $C_1$ to $C_n$, occurring at any time before measurement of these central qubits.
If less than $n/2$ errors occur then the encoded one-qubit state is perfectly teleported across the cluster, from A to B.
\textbf{b}.
Conceptually equivalent quantum logic circuit.
}
\label{levelscheme}
\vspace{4mm}
\end{figure*}

\begin{figure*}[t]
\vspace{5mm}
\includegraphics[width=1.5 \columnwidth]{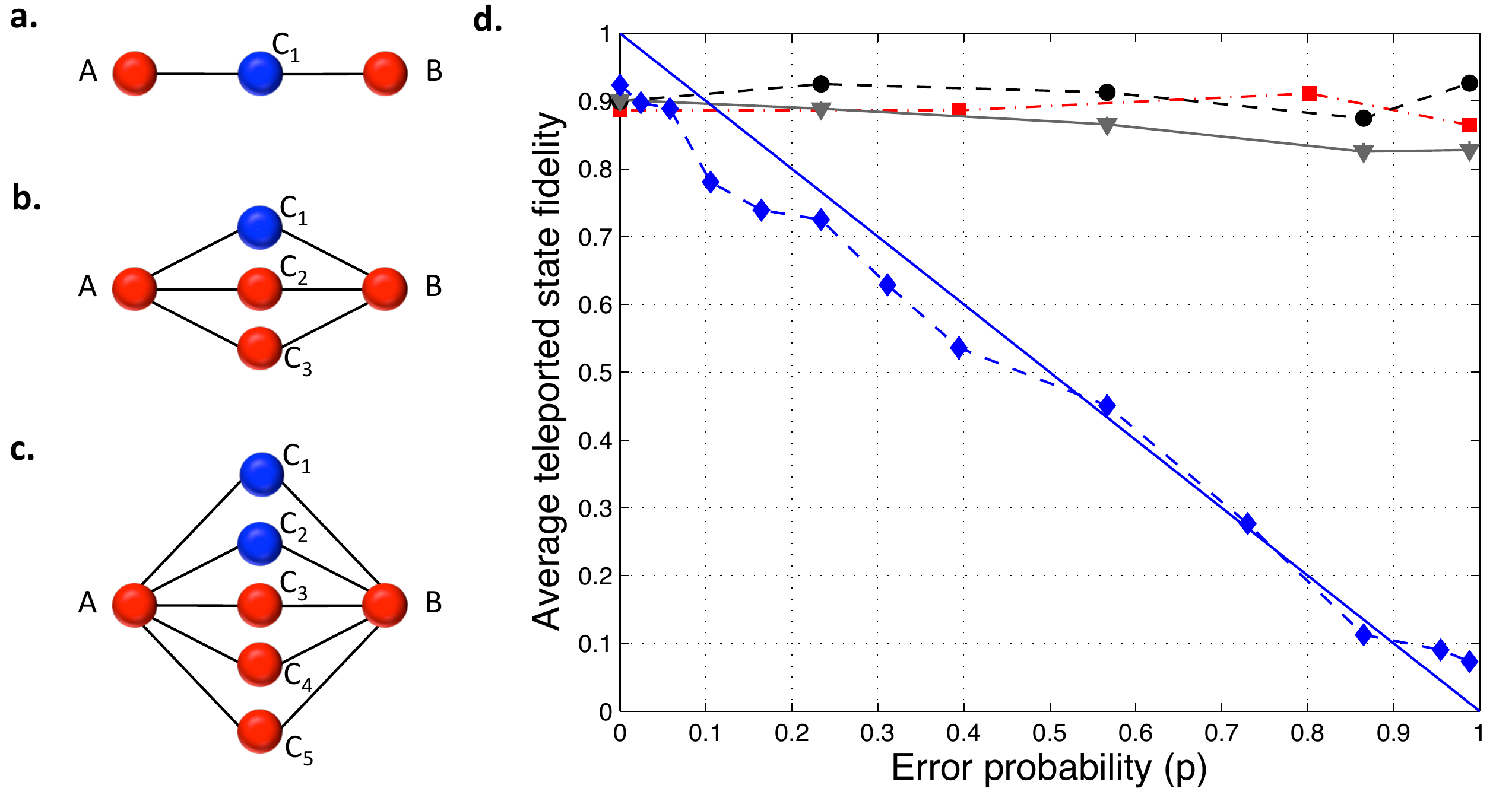}
\vspace{5mm}
\caption{\label{error} \textbf{Quantum error correction performance against errors on subsets of codeword qubits}.
\textbf{a} to \textbf{c}. Graph states $\ket{EC_n}$  for  $n=$ 1, 3 and 5, respectively. 
Errors are applied to qubits in blue. 
\textbf{d} Solid blue line: ideal case in \textbf{a}. Experimental results for cases \textbf{a} - \textbf{c} are shown as blue diamonds, red squares and grey inverted triangles (two errors), respectively. Black circles show case \textbf{c} for only a single error applied to $C_1$. For more details see supplementary material. 
Errors are one standard deviation and derived from quantum projection noise. 
}
\label{levelscheme}
\vspace{4mm}
\end{figure*}

\begin{figure*}[t]
\vspace{-5mm}
\includegraphics[width=1.5 \columnwidth]{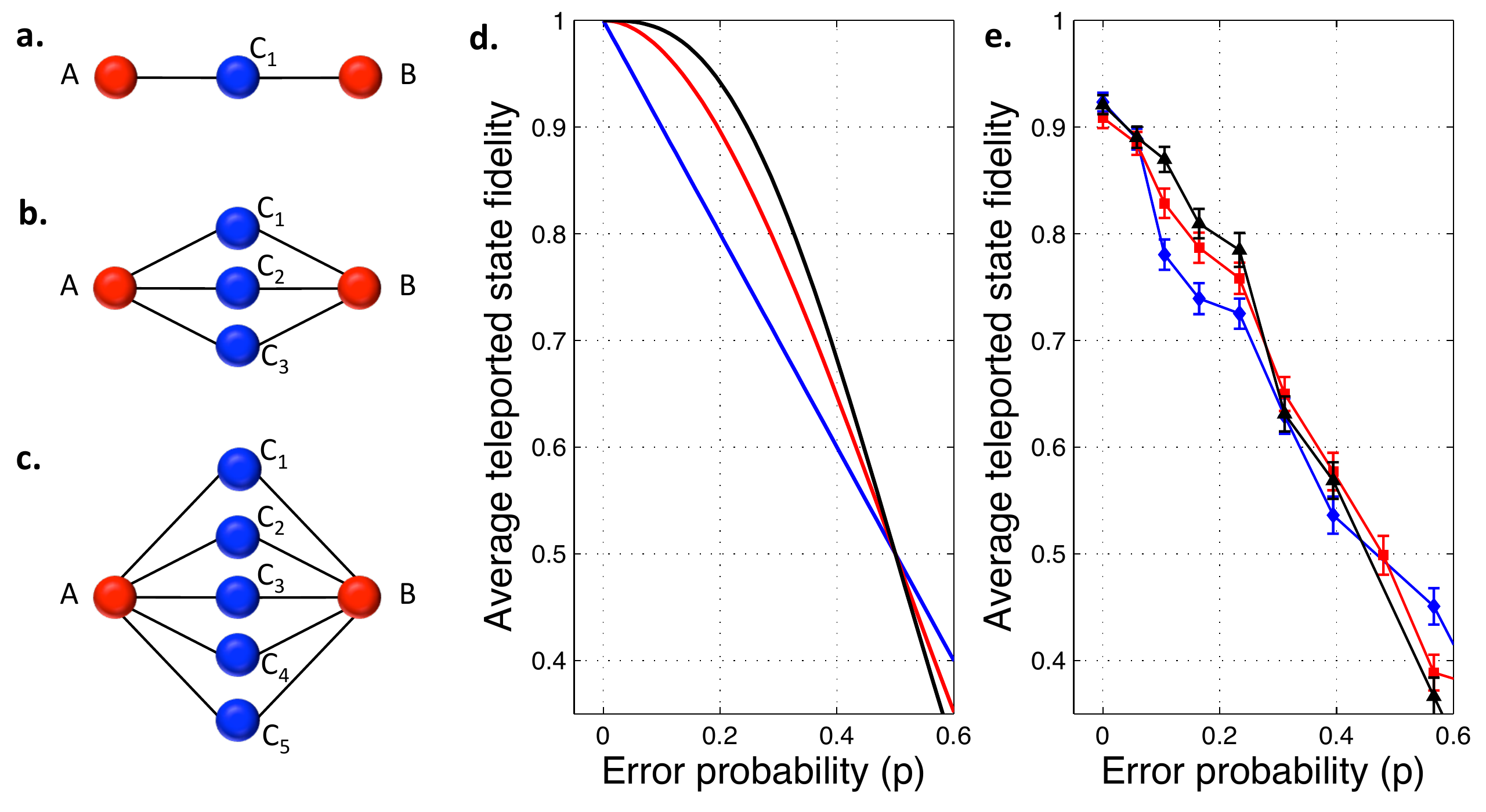}
\vspace{0mm}
\caption{\label{error} \textbf{Quantum error correction performance against errors on all codeword qubits}.
\textbf{a} - \textbf{c}. Graph states $\ket{EC_n}$  for  $n=$ 1, 3 and 5, respectively. 
Errors are applied to qubits in blue. 
\textbf{d} Ideal performance for cases \textbf{a} - \textbf{c} shown as solid blue red and black lines, respectively. Increasing the codeword length ($n$) improves performance for $p<0.5$.
\textbf{e} Experimental results for cases \textbf{a} - \textbf{c} shown as blue diamonds, red squares and black triangles, respectively. Errors are one standard deviation and derived from quantum projection noise. 
}
\label{levelscheme}
\vspace{4mm}
\end{figure*}

\begin{table}[htdp]
\caption{
\label{table} \textbf{Properties of experimentally generated graph states}. 
%For details on the measures presented, see supplementary material. 
Fidelity and purity derived from the tomographically reconstructed density state ($\rho$). $F{=}\Tr{[\rho\ket{\psi}\bra{\psi}]}$, $P{=}\Tr{[\rho^2]}$, where $\ket{\psi}$ is the ideal state. $\cal{B}$$_n$ is equivalent to the state fidelity, derived from a subset of tomographic measurements. Values on the rhs of the inequality are the maximum allowed by LHV models (see supplementary material). NM: not measured. Errors are one standard deviation and derived from quantum projection noise. 
}
\begin{center}
\begin{tabular}{|c|c|c|c|c|}
Cluster & qubits & Fidelity ($F$) & Purity ($P$) & Multipartite Bell inequality ${\cal{B}}_n$\\
\hline\hline
$LC_4$  & 4 & 0.841$\pm{0.006}$ & 0.82$\pm{0.01}$ & 0.85$\pm{0.02}> 0.75$\\
$RC_4$ & 4 & 0.847$\pm{0.007}$ & 0.75$\pm{0.01}$ & 0.86$\pm{0.02}> 0.75$\\
$EC_1$ & 3 & 0.920$\pm{0.005}$ & 0.88$\pm{0.01}$ & 0.92$\pm{0.02}> 0.75$\\
$EC_3$ & 5 & 0.843$\pm{0.005}$ &  0.787$\pm{0.008}$ & 0.86$\pm{0.01}> 0.75$\\
$EC_5$ & 7 & NM & NM & 0.73$\pm{0.01}>0.625$\\
\hline
\end{tabular}
\end{center}
\label{stats}
\end{table}%

\clearpage

%%%%%%%%%
%%%%%%%%%
%%%%%%%%%

\begin{widetext}
\noindent \Large{\textbf{Supplementary Material: \\
Measurement-based quantum computation with trapped ions}}
\\
\end{widetext}

\setcounter{equation}{0}

\section{General experimental details}

We use a 3D linear Paul blade trap built with nearest ion-electrode separation of $560\mu m$. Depending on the number of ions in the graph state to be generated, we chose trapping parameters (frequencies) that were slightly different, but were approximately $1$, $3.4$ and $3.5$ MHz for the axial and radial centre-of-mass vibrational modes. An important feature is that the trap has holes in the axial end-cap electrodes, allowing direct laser access along the principle axis of the ion string.  

We use  $^{40}$Ca$^{+}$ ions. Two electronic Zeeman states encode a qubit ($|D_{5/2},m=+3/2\rangle{=}\ket{0}$, $|S_{1/2},m=+1/2\rangle{=}\ket{1}$)  which are coupled by an electric quadrupole transition at 729~nm. Each experiment begins with Doppler cooling (3.5$m$s) and optical pumping  (20$\mu$s) all ions in the string into the $\ket{1}$ state. Next both the axial COM  (4.5$m$s) and stretch modes  (4.5$m$s) are subsequently ground-state cooled via sideband cooling. A final frequency-resolved optical pumping step ensures that all ions are in the $\ket{1}$ state with high probability (500$\mu$s). 
Qubit state detection is done via electron shelving: light at 397~nm and 866~nm exciting the $S_{1/2}$-$P_{1/2}$ and $D_{3/2}$-$P_{1/2}$ transitions, respectively, is sent in to the ion string, which will only scatter if the electron is found to be in state $\ket{1}$. Scattered light at 397~nm is detected using both a photo-multiplier tube and a CCD camera. The detection time is $5m$s.

\section{Graph-state generation tools}

Beginning with the $n$ ionic-qubit state $\ket{1}^{\otimes n}$, we manipulate the quantum state using laser beams at 729~nm. In the main text we refer to three distinct tools that are employed to generate graph-states. We now explain these tools in more detail:

\subsection{Qubit-qubit interaction}

An effective long-range qubit-qubit interaction with a Hamiltonian of the form $H\propto \sum_{a<b}X_aX_b$ can be turned on for arbitrary times. When applied to ionic-qubits in $\ket{1}^{\otimes n}$ this periodically generates fully connected graph states that are equivalent to GHZ states \cite{PhysRevLett.106.130506, PhysRevLett.82.1835}. The interaction is `effective' since it is mediated by one of the collective vibrational modes of the ion string. Only at certain times  during the dynamics can this additional degree of freedom be ignored, and at these times the unitary dynamics is equivalent to turning on a qubit-qubit interaction for a particular time. 

Specifically, we use the M{\o}lmer-S{\o}rensen interaction \cite{PhysRevLett.82.1971}: a laser beam is sent along the axial direction of the ion string, with two frequency components symmetrically detuned by $\pm(\nu+\delta)$ from of the axial centre-of-mass (COM) vibrational mode sidebands of the qubit transition. This beam couples to all ions equally. Here $\nu$ is the angular frequency of the COM mode. At integer multiples of time $t{=}\frac{2\pi}{\delta}$ the motional state will come back to rest and the ionic-qubits will pick up a geometric phases which depend nonlinearly on the state of the qubits. Employing this interaction for an $n$ ionic-qubit string, for a time $t{=}\frac{2\pi}{\delta}$ realizes the unitary operation

\begin{equation}
U^{X}_{ms}(\theta)=\exp[-i\theta \sum_{a{<}b}X_aX_b]
\label{MS}
\end{equation}

where $\theta{=}\pi\eta^2\Omega^2/\delta^2$, $\eta$ is the n-ion Lamb-Dicke parameter and $\Omega$ is the Rabi frequency. For a single $^{40}Ca^{+}$ ion at a COM frequency of 1MHz, $\eta_1=0.097$. For $n$ ions $\eta=\eta_1/\sqrt{n}$. When applied to the initial state $\ket{1}^{\otimes n}$, the operation $U_{ms}^X(\pi/4)$ generates GHZ states, i.e. $U_{ms}^X(\pi/4)\ket{1}^{\otimes n}=(\ket{1}^{\otimes n}-i\ket{0}^{\otimes n})/\sqrt{2}$. We typically used $\delta=2\pi\times20$ KHz, although the value varied by up to a factor of two in the pulse sequences used to generate different graph states.

\subsection{Single-qubit phase flips}

We realise single qubit phase-shift gates described by the unitary

\begin{equation}
U_z^{k}(\theta)=\exp[-i\frac{\theta}{2}Z_k]
\label{AC}
\end{equation}

which applies a Pauli $Z$ rotation on ionic-qubit $k$, by angle $\theta$, using a tightly focused far off-resonant laser beam interacting only with the $k$-th ion, via the AC-Stark effect. This beam comes in perpendicular to the string and can be switched between any ionic-qubit in the string over timescales of approximately $10\mu$s using an acousto-optic deflector. 

The operations in equations (1) and (2) can be combined to generate any arbitrary graph state. The proof of this is done by showing that (1) and (2) can be combined to generated a maximally entangling gate between any two ionic-qubits in a string, i.e. the operation $\exp[-iX_aX_b\pi/4]$. This proof is given in \cite{PhysRevA.79.012312}. This is sufficient because any graph state of qubits can be built by starting with a simple product state and applying a sequence of two-qubit maximally entangling gates. 

\subsection{Hiding pulses}

\begin{figure}[tb]
\centering
\includegraphics[width=1\columnwidth]{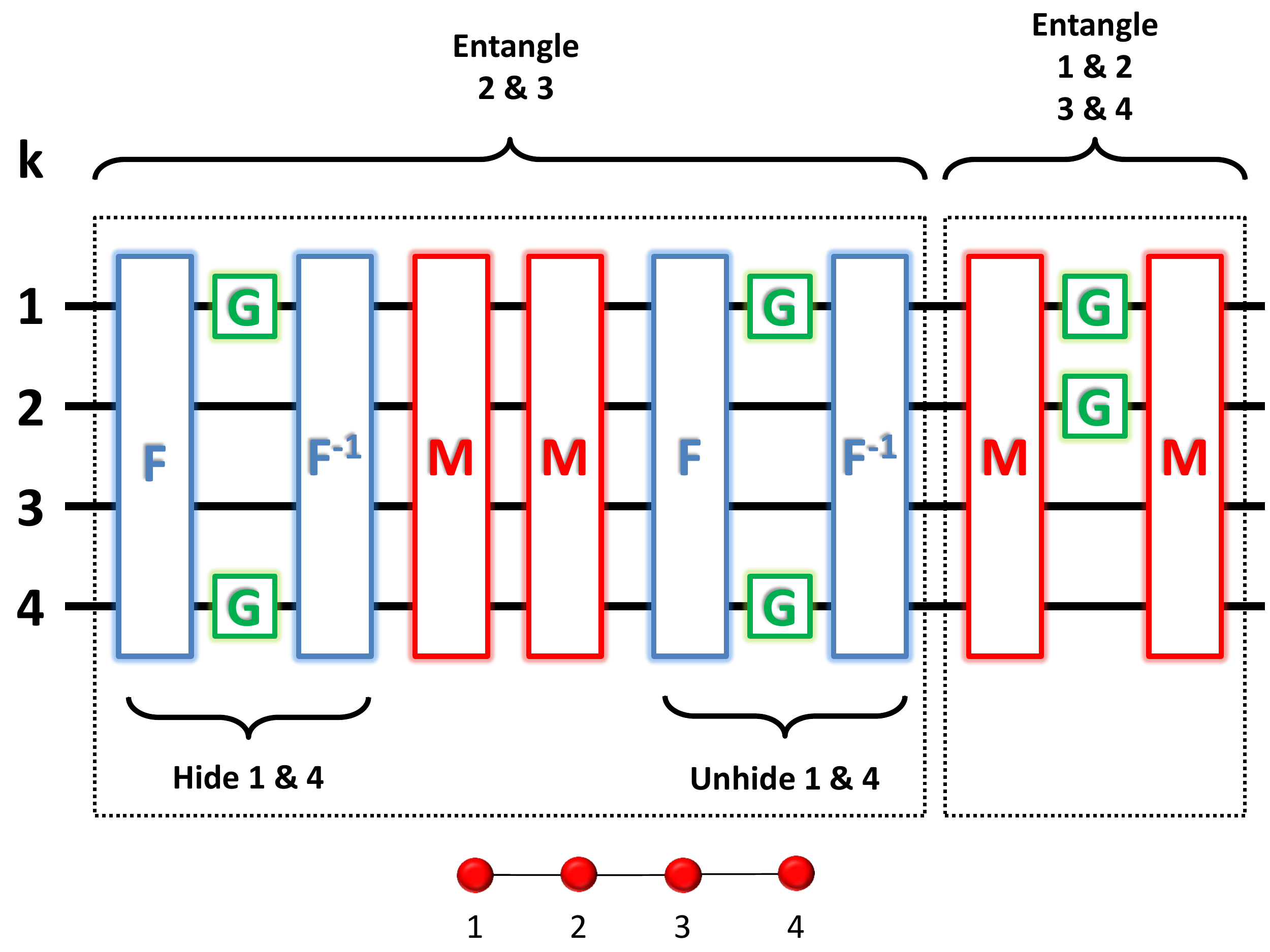}
\caption{\textbf{Experimental laser-pulse (logic-gate) sequence used to generate the 4-qubit linear cluster $\ket{E_{LC4}}$}. 
The initial state is $\ket{1}^{\otimes n}$. 
With reference to equations \ref{MS} and \ref{AC}, the operations are $G=U_z^{k}(\pi)$ and $M=U_{ms}^X(\pi/8)$. The operations $F=\exp[-i\frac{\pi}{4}\sum_k \hat{X}_k]$ and $F^{-1}$ are hiding 729~nm $\pi$-pulses done with on the $\ket{1} - |D_{5/2},m=5/2\rangle$ transition. The figure is divided into two distinct parts. The first part entangles ionic-qubits 2 and 3, by temporarily hiding ion-qubits 1 and 2 in the $|D_{5/2},m=5/2\rangle$ state and using entangling operations $M$. The second part, when applied alone, would entangle only  ion pairs 1 \& 2 and 3 \& 4 using refocusing pulses $G$. The state generated, $\ket{E_{LC4}}$, defined eq.~(\ref{E_LC_4}), can be converted to the standard cluster built with CPHASE gates $\ket{LC_4}$ by the local unitary operations in table \ref{tablecorlin}. 
}
\label{gates_LC4}
\end{figure}

While the previous two tools are sufficient to generate any graph-state in principle, in practice more efficient methods are possible by adding other tools, such as hiding pulses. Hiding pulses exploit other $S_{1/2}$ and $D_{5/2}$ electronic Zeeman states in addition to those that we have chosen to encode a qubit. Transitions between these other levels are many MHz different than the qubit transition and can therefore be used to temporarily hide quantum states from subsequent frequency-selective operations.  The most general case, where a qubit in an  unknown state can be hidden, is described in \cite{Barreiro:2011fk}. We use a simpler version as part of the experimental sequence to generate the linear cluster $\ket{LC_4}$ (see figure \ref{gates_LC4}), which exploits the fact that the states of the qubits to be hidden are known and equal to $\ket{1}$. In this case the electron can simply be transferred in and out of one of the $D_{5/2}$ Zeeman states (other than $|D_{5/2},m=+3/2\rangle=\ket{1}$) with a single 729nm pulse.

\section{Data processing}\label{Data_processing}

The methods that we employ to generate graph states experimentally are equivalent to preparing all $n$ qubits in the graph in $\ket{1}^{\otimes n}$, then applying pairwise entangling operations $\exp[-i\frac{\pi}{4}X_aX_b]$ between all qubits pairs (a,b) connected by a vertex. This is equivalent to the more well known process, of applying CPHASE gates to qubits prepared in $\ket{+}^{\otimes n}$, where $\ket{+}=(\ket{0}+\ket{1})/\sqrt{2}$, up to single-qubit unitary operations. The operations, which have to be applied to the experimental graph states to convert them to the `CPHASE graph states' are presented in tables \ref{tablecorlin}, \ref{rep7cor} and \ref{tablecorbox}. Note that all of these operations map Pauli operators to other Pauli operators. Consequently, measuring the expectation value of any product of Pauli operators on the CPHASE states is equivalent to measuring the expectation value of another product of Pauli operators on the experimentally generated states.  For example, measuring qubits 1,2 3 and 4 of the 4-qubit linear cluster $\ket{LC_4}$ in the Z, Z, X, X basis', respectively, is equivalent to measuring the same qubits in the experimental version $\ket{E_{LC_4}}$ of the $\ket{LC_4}$ in the Y, X, Z and Z basis', respectively (See table \ref{tablecorlin}). We simply reinterpret our measurement results in this way to accommodate these correction operations. In conclusion, the first step in the data processing is to reinterpret the measurement basis. 
Note that there is no fundamental difference between states generated in either way, we just choose to present all our results with respect to graph states built in the more common way using CPHASE gates.

\begin{table}[tb]
\caption{Correction operations that have to be applied to the experimentally generated linear cluster state $\ket{E_{LC4}}$ to convert it to the standard linear cluster state $\ket{LC4}$. $H$ denotes the Hadamard operation.
}
\label{tablecorlin}
\centering
\begin{tabular}{c c}
\hline \hline
qubit $\# 1$ & $H Z e^{-i\pi/4X}$\\
qubit $\# 2$ & $H ZX$\\
qubit $\# 3$ & $H X$\\
qubit $\# 4$ & $H e^{-i\pi/4X}$\\
\hline
\end{tabular}
\end{table}

\begin{table}[tb]
\caption{Correction operations that have to be applied to the experimentally generated (n+2)-qubit graph state for quantum error correction $\ket{E_{ECn}}$, to convert it into the standard form $\ket{EC_n}$.}
\label{rep7cor}
\centering
\begin{tabular}{c c}
\hline \hline
qubit $A$ & $He^{i\pi/4X}Z$\\
qubit $C_1$ to $C_n$ & $H$\\
qubit $ B$ & $He^{i\pi/4X}Z$\\
\hline
\end{tabular}
\end{table}

\begin{table}[t]
\caption{Correction operations that have to be applied to the experimentally generated ring cluster state $\ket{E_{RC4}}$, to convert it into the standard form $\ket{RC4}$.}
\label{tablecorbox}
\centering
\begin{tabular}{c c}
\hline \hline
qubit $\# 1$ & $H X Z$\\
qubit $\# 2$ & $H X$\\
qubit $\# 3$ & $H X$\\
qubit $\# 4$ & $H X Z$\\
\hline
\end{tabular}
\end{table}

%e.g. for a one-qubit output state, $q=1$, we measure 6 probabilities

%We characterise each of these states by measuring expectation values of observables built from all $3^n$ combinations of products of Pauli operators. 

The next data processing steps are to simulate the action of perfect feedforward. We now summarise how this is done. More details about each instance are given in the remaining sections of this supplementary material. 

Feedforward operations, which are always equivalent to single-qubit rotations, can be divided into two kinds. Firstly there are the kind that can be commuted all the way through a measurement-based protocol to act only on the final output state - these final `bi-product operators' can depend on the outcomes of previous measurements, but do not change the basis in which measurements have to be made en-route to the output state. The two-qubit gate and error correction demonstrations, presented in the main text, all require feedforward only of this kind: byproduct operators applied to the final states which depend on the previous measurement outcomes. 

We always deal with these by-product operators in the following way: for a $(p+q)$-qubit graph state, with $q$ output qubits, there are $2^p$ possible outcomes for a given measurement configuration of the $p$ qubits. Correspondingly there are $2^p$, potentially different, output states of the $q$ qubits, all ideally related by bi-product operators (local unitaries).  We perform enough measurements to fully characterise each of the $2^p$ different $q$-qubit output states. Specifically we estimate the probabilities of observing all $2^q$ eigenstates of all $3^q$ observables built from combinations of products of Pauli operators. Each probability is therefore associated with an eigenstate. 
The bi-product operators are included by reassigning this association i.e. changing the eigenstates (the Heisenberg picture). 
After this, a single set of $2^q \times 3^q$ probabilities is established by summing up all $2^p$ instances (one for each output state), and the output density matrix is reconstructed via maximum-likelihood tomography.  Note that another approach could be to reconstruct all $2^p$ output states via maximum-likelihood tomography, apply the bi-product operators and then take the sum weighted by the probabilities of observing each state. However, this is impractical since the probabilities of certain outcomes can be vanishingly small thereby introducing huge uncertainties in the state reconstructions.  Our approach, of summing probabilities and perform state reconstruction once at the end is exactly equivalent to the action of perfect feedforward followed by state reconstruction. 

The second kind of feedforward is that which changes the basis of subsequent measurements. This kind is only relevant to our demonstration of one-qubit gates using the 4-qubit linear cluster $\ket{LC_4}$, where several qubits are measured sequentially before the output. We now describe our approach. A circuit consisting of one-qubit rotations can be simulated by measuring the 4-qubit linear cluster $\ket{LC_4}$ according to the measurement pattern shown in figure 1 of the main text. The angles of the rotations are determined by three parameters $\alpha$, $\beta$ and $\gamma$, respectively. The outcome of the measurement $B(\alpha)$, which projects qubit 1 onto the states $|\pm\alpha\rangle=(|0\rangle\pm e^{i\alpha}|1\rangle)/\sqrt{2}$, determines the sign of the measurement basis of qubits 2 and 3 i.e $B(\pm{\beta})$ and $B(\pm{\gamma})$, respectively. (How this decision is made is given in a later section of this text). Consequently there are four possible measurement basis combinations, each with $2^3$ possible outcomes (eigenvalues). Our approach is to measure an over-complete set of expectation values, required to characterise a one-qubit state, for all $4\times2^3$ outcomes. Each set has a bi-product operator which is included in the way described in the previous section, resulting in a single set of expectation values that is used to reconstruct the one-qubit output state---examples of which are presented on the Bloch sphere (figure 1d, main text). Note that this whole process is simplified by choosing $\beta$ and $\alpha$ from the set $[0,\pm{\pi/2}]$, such that $B(\pm{\alpha})$ and $B(\pm{\beta})$ are the same observable up to a global phase. 

\section{Demonstration of a measurement-based universal gate set}

This section provides more details on our demonstration of a universal gate set in the MBQC model. The main results from this are presented in figures 1 and 2 in main text, where one- and two-qubits gates are demonstrated, respectively.

\subsection{Resource state: 4 qubit linear cluster}

The 4 qubit linear cluster state $\ket{LC_4}$, shown at the bottom of figure \ref{gates_LC4},  can be generated by preparing all 4 qubits in $\ket{+}$ and applying CPHASE gates between every pair of qubits connected by an edge. i.e. $\ket{LC_4}=C_{34}C_{23}C_{12}\ket{+}^{\otimes n}$. Here $C_{ij}$ is a CPHASE between qubits $i$ and $j$. We experimentally generate the state $\ket{E_{LC_4}}$,

\begin{eqnarray}
\label{E_LC_4}
\ket{E_{LC_4}}&=&\frac{1}{\sqrt{8}}(\ket{0000}-i\ket{0011}-\ket{0101}\ldots\\
&&\phantom{ii}-i\ket{0110}+i\ket{1001}-\ket{1010}\ldots\nonumber\\
&&\phantom{ii}-i\ket{1100}-\ket{1111})\nonumber
\end{eqnarray}

which is equivalent to $\ket{LC_4}$ up to single-qubit unitary operations (i.e. they are locally equivalent). The local unitary operators, which have to be applied to $\ket{E_{LC_4}}$ to convert it to $\ket{LC_4}$ are shown in table \ref{tablecorlin}. These local unitaries do not change the non-local properties of the clusters and the method we employ to deal with them is explained in section \ref{Data_processing}.

The experimental laser pulse sequence used to generate $\ket{E_{LC4}}$ consists of two parts and is presented in figure $\ref{gates_LC4}$. The first part entangles ionic-qubits 2 and 3 out of a four ion string. We now describe this part. Beginning with $\ket{1111}$, hiding pulses move the $|S_{1/2},m{=}{+}1/2\rangle{=}\ket{1}$ electrons of ions 1 and 4 into the $|D_{5/2},m{=}{+}5/2\rangle$ states. The frequency selective operation $U^{X}_{ms}(\pi/4)$ then fully entangles only the remaining qubits 2 and 3. Unhiding pulses then restore both ionic-qubits 1 and 4 into $\ket{1}$.  At this point the four-qubit state is $\ket{111}-i\ket{1001})/\sqrt{2}$. The second part generates entanglement between qubits 1 and 2, and between qubits 3 and 4, simultaneously. The approach is to use refocusing $U^{k}_z(\pi)$ pulses (Eqn. (2)) between $U^{X}_{ms}(\pi/8)$ operations to remove particular interactions as proposed in \cite{PhysRevA.79.012312}. Examples of this refocusing technique are shown in figure \ref{gates_LC4} (in the last group of laser pulses where pairs of qubits are entangled) and figure \ref{gates_EC} (in the first group of pulses entitled `cluster generation').

\subsection{1 qubit gates}

By measuring $\ket{LC_4}$ in the order presented in figure 1 of the main text, a quantum circuit is simulated that consists of a series of one-qubit rotations. The angles of rotation $\alpha$, $\beta$ and $\gamma$ are determined by the choice of measurement basis, as described in the main text. The bases in which qubits 2 and 3 are measured have to be adapted to the previous measurement outcomes. Let $s=0$ denote the +1 outcome and $s=1$ the -1 outcome of a measurement. Then one has to choose the measurement angles of qubit 2 and 3 as $(-1)^{s_1}\beta$ and $(-1)^{s_2}\gamma$, respectively. The subscript denotes the measured qubit. Finally, there is a byproduct operator on the final qubit 4 given by $X^{s_1+s_3}Z^{s_2}$.

\subsection{2 qubit gates}

By measuring $\ket{LC_4}$ in the order presented in figure 2 of the main text, one simulates a quantum circuit that consists of both one- and two-qubit gates. The angles of rotation are determined by the choice of measurement basis, as described in the main text. 
The final byproduct operators for the circuit are given by $X^{s_1}Z^{s_4}$ for qubit 2 and $X^{s_4}Z^{s_1}$ for qubit 3. There is no feedforward since there are only input and output qubits and no intermediate measurements.

\section{Demonstration of measurement-based quantum error correction}

This section provides more details on our demonstration of measurement-based quantum error correction, the main results of which are presented in figures 4 and 5 in the main text. 

\subsection{Resource graph-state}

When generated using CPHASE gates, the $(n+2)$-qubit resource state for our quantum error correction demonstration is given by
\begin{equation}
2\ket{EC_n}{=}\left(\ket{0}_A\ket{+}^{\otimes n}+\ket{1}_A\ket{-}^{\otimes n}\right)\ket{0}_B+\left(\ket{0}_A\ket{-}^{\otimes n}+\ket{1}_A\ket{+}^{\otimes n}\right)\ket{1}_B
\end{equation}
Experimentally we generate the locally-equivalent state $\ket{E_{ECn}}$, where
\begin{equation}
2\ket{E_{ECn}}{=}\left(-i\ket{-}_A\ket{0}^{\otimes n}+\ket{+}_A\ket{1}^{\otimes n}\right)\ket{-}_B+
\left(\ket{-}_A\ket{1}^{\otimes n}+i\ket{+}_A\ket{0}^{\otimes n}\right)\ket{+}_B
\end{equation}

 The local unitary operators, which have to be applied to $\ket{E_{ECn}}$ to convert it to $\ket{EC_n}$ are shown in table \ref{rep7cor}. 
These operators are taken into account in the way described in the section \ref{Data_processing}. Because of these corrections, in the laboratory, our graph states protect against $X$ errors (bit flip errors). 

\begin{figure}[tb]
\centering
\includegraphics[width=1 \columnwidth]{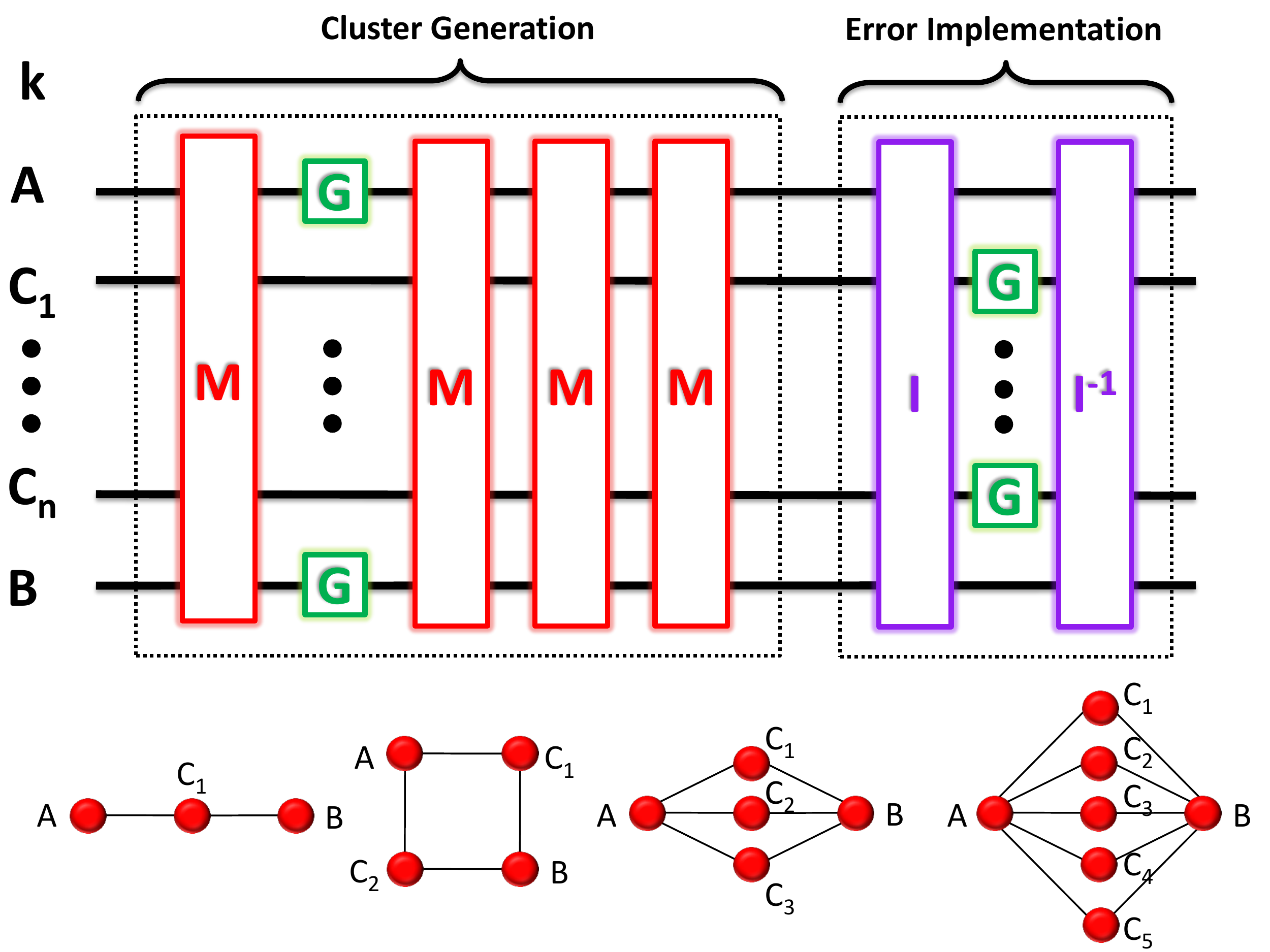}
\caption{\textbf{Experimental laser-pulse (logic-gate) sequence used to generate the ($n{+}2$)-qubit graph states $\ket{E_{ECn}}$}. 
The initial state is $\ket{1}^{\otimes n}$. 
With reference to equations \ref{MS} and \ref{AC}, the operations are $G=U_z^{k}(\pi)$ and $M=U_{ms}^X(\pi/8)$. After the sequence of pulses labelled `Cluster generation' the state $\ket{E_{ECn}}$ is prepared, which can be converted to the standard graph built with CPHASE gates $\ket{EC_{n}}$ by the local unitary operations in table \ref{rep7cor}. 
The operations $I=\exp[-i\frac{\theta}{2}\sum_kX_k]$ and $I^{-1}$ are done with 729~nm pulses on the qubit transition. 
The sequence of pulses labelled `Error Implementation' is equivalent to performing an operation $\exp[i\theta X]$ on all those ionic-qubits in the set $C_n$ that receive a $G$ operation between the two $I$ pulses: After measurement of these qubits in the $Z$ basis, these $X$ rotations are projected into $X$-flip errors occurring with probability $p=sin^{2}(\theta/2)$. After correcting for the operations in table \ref{rep7cor}, $X$ errors on $\ket{E_{ECn}}$ are converted to $Z$ errors on $\ket{EC_n}$. All the graph states generated using this pulse sequence are shown at the bottom. From left to right they are the cases $n=$[1,2,3,5].
}
\label{gates_EC}
\end{figure}

The experimental laser-pulse sequence used to generate  $\ket{E_{ECn}}$ (for $n{=}1,2,3,5$) and to controllably implement errors, is presented in figure \ref{gates_EC}. Note that the $n{=}2$ case is the 4-qubit ring cluster $\ket{RC4}$.

\subsection{Detailed example of the quantum error correction protocol for 5 qubit graph state: $\ket{EC_3}$}

The 5-qubit graph state $\ket{EC_{3}}$, used to demonstrate the measurement-based three-qubit repetition code, is shown at the bottom of figure \ref{gates_EC} and is an example of equation 4 for $n{=}3$. After preparation of this graph state, the protocol proceeds as follows:
\\
1. A 1-qubit state $\ket{\psi}$, or the orthogonal state, is encoded by measuring qubit A in a basis where these are the eigenstates.  The effect is to distribute either state non-locally amongst the remaining 4 qubits. 
\\
2. Unitary phase rotations (Z)  are controllably applied to all, or a subset of, the central codeword qubits $C_1, C_2, C_3$. After step 3 these will become discretised phase flip errors. 
\\
3. Each of the 3 central codeword qubits is measured in the X basis, yielding one of $8$ possible outcomes. This simultaneously decodes the state and reveals if up to one of the central qubits has experienced a phase flip (i.e. determines the error syndrome). 
\\
4. A 1-qubit recovery operation, determined by the outcome in step 3 is applied to the output state, stored in qubit $B$. This recovers the initial encoded 1-qubit state.  \\
\\
The recovery operators in step 4 are shown in table \ref{recovery}. We call these recovery operators, rather than error corrections, since they can be necessary even if no error has occurred. Generalization to the five qubit repetition code is straightforward: whenever the majority of measurement outcomes is +, the recovery operator is $\mathbb{I}$, otherwise it is $Z$.

Note that the outcome of the measurement in step 1 is unknown \textit{a priori} and so either $\ket{\psi}$ or its orthogonal state $\ket{\tilde{\psi}}$ is encoded into the cluster, protected and eventually read out. In the cases where $\ket{\psi}$ and $\ket{\tilde{\psi}}$ are related by a single Pauli operator then the output $\ket{\psi}$ can be deterministically recovered via feedforward. If one wished to deterministically encode any state, even an unknown state, then this is possible by introducing another qubit prepared in this state, performing a Bell-state measurement between this qubit and qubit A, and performing feedforward dependant on which of the four outcomes are observed.

\begin{table}[t]
\caption{Recovery operations for the possible outcomes of measuring the three codeword qubits, in the Pauli X basis, in our QEC demonstration using the 5-qubit graph state $\ket{EC_3}$ .}
\label{recovery}
\centering
\begin{tabular}{c c c c c c c c c}
\hline \hline
qubit $C_1$ & $+$ & $+$ & $+$ & $+$ & $-$ & $-$ & $-$ & $-$\\
qubit $C_2$ & $+$ & $+$ & $-$ & $-$ & $+$ & $+$ & $-$ & $-$\\
qubit $C_3$ & $+$ & $-$ & $+$ & $-$ & $+$ & $-$ & $+$ & $-$\\
recovery op. & $\mathbb{I}$ & $\mathbb{I}$ & $\mathbb{I}$ & $Z$ & $\mathbb{I}$ & $Z$ & $Z$ & $Z$\\
\hline
\end{tabular}
\vspace{5mm}
\end{table}

\subsection{Additional experimental results, discussion and analysis}

\begin{figure}[tb]
\centering
\includegraphics[width=1 \columnwidth]{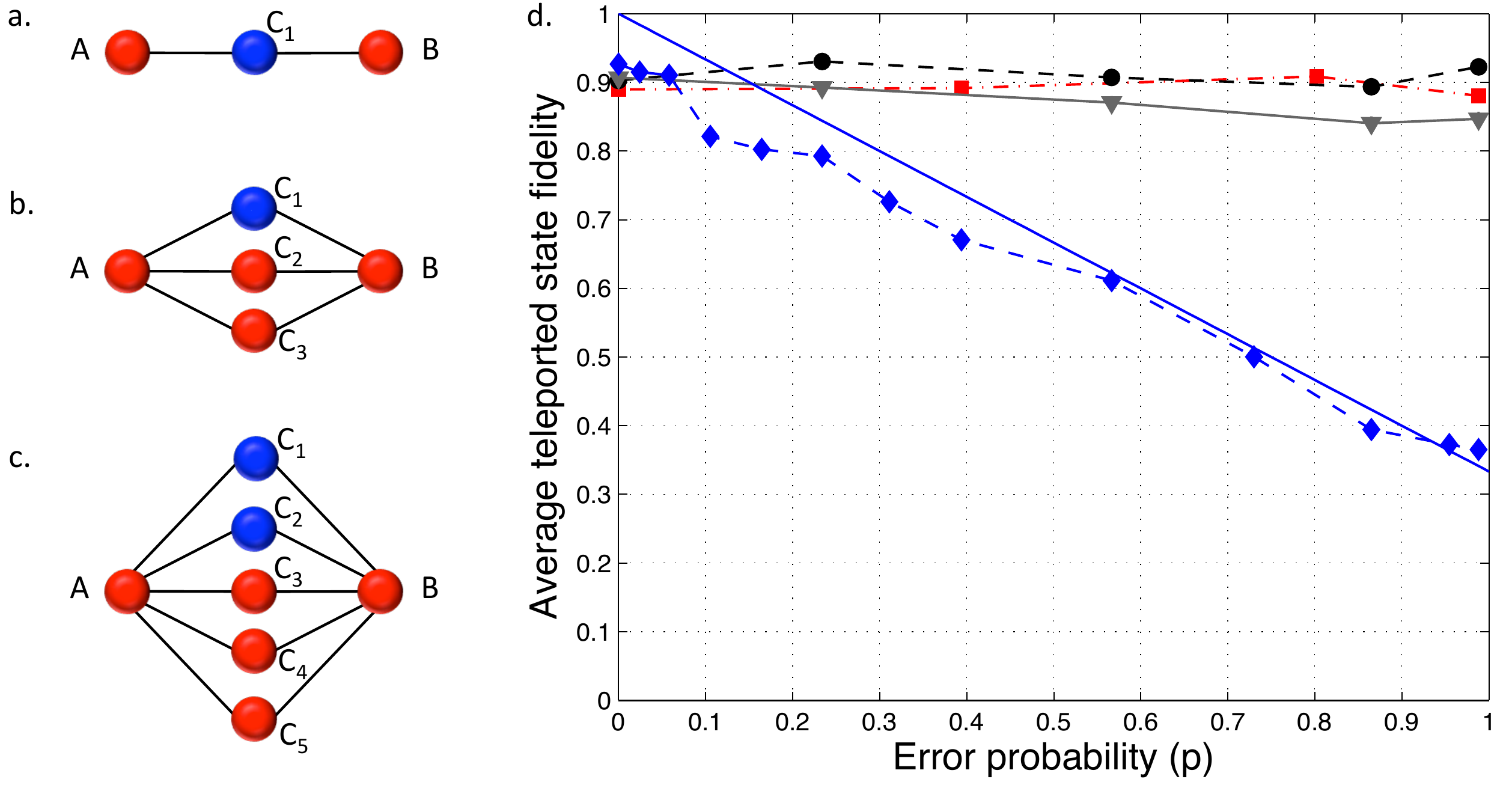}
\caption{ \textbf{Quantum error correction performance against localised errors, averaged over 6 input states}.
This figure is a complement to figure 4 in the main text. It presents results from the same experiment but differs in that the average teleportation fidelity is now averaged of 6 input states (compared to 4 previously). The two new input states are eigenstates of the error process ($Z$ errors). See text for more details. 
\textbf{a} to \textbf{c}. Clusters $\ket{EC_n}$ for  $n=$ 1, 3 and 5, respectively. Rotations $R_z=\exp(-i\frac{\theta}{2}Z)$ are applied to the qubits in blue, which after measurement are projected into Z flip errors with probability $p{=}\sin^2(\theta)$. \textbf{d} Experimental results. Solid blue line: ideal case in \textbf{a}. Experimental results for cases \textbf{a} - \textbf{c} are shown as blue diamonds, red squares and grey inverted triangles (two errors), respectively. Black circles show case \textbf{c} for only a single error applied to $C_1$.
}
\label{all6_local}
\end{figure}

\begin{figure}[tb]
\centering
\includegraphics[width=1 \columnwidth]{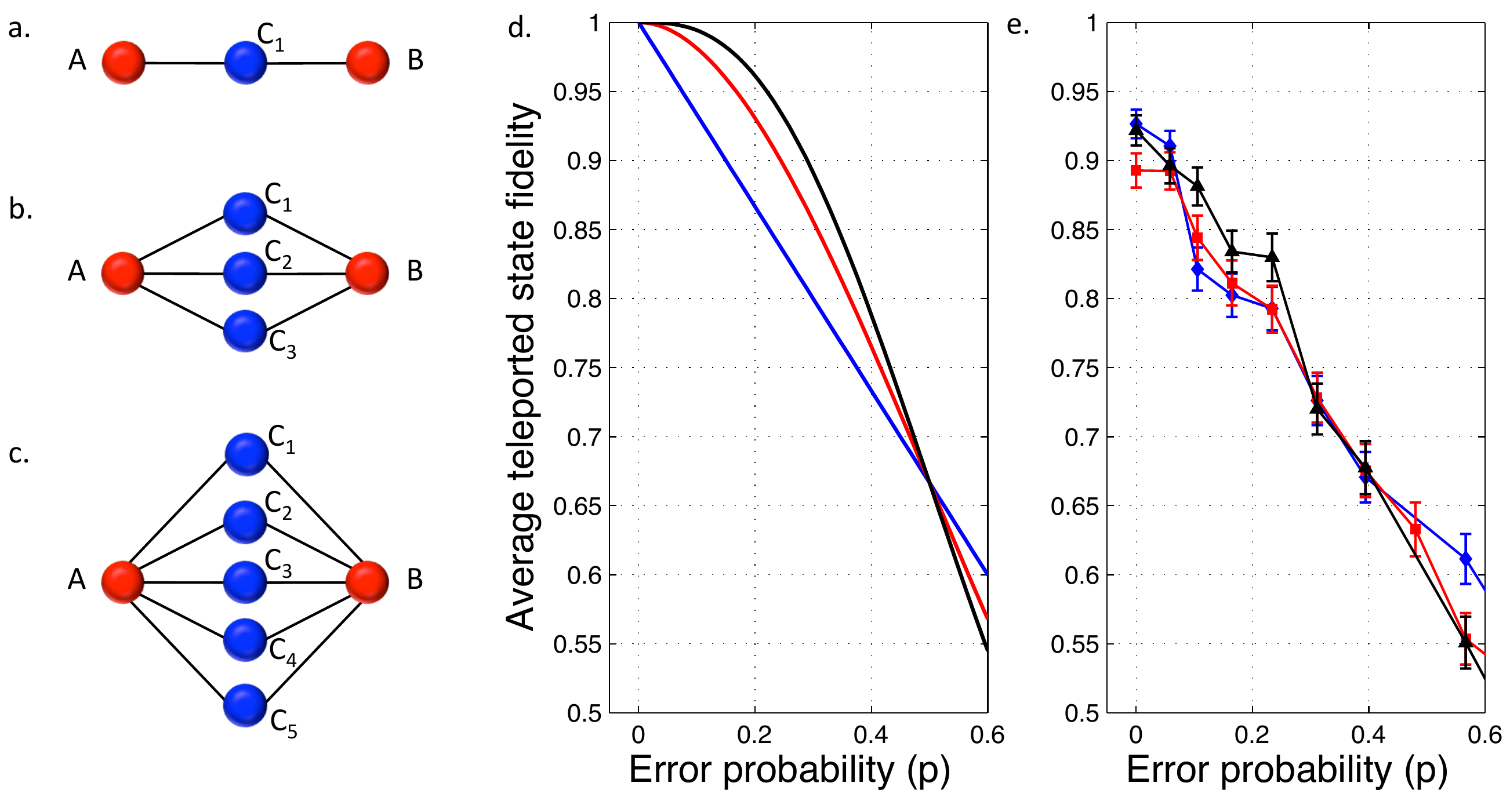}
\caption{\textbf{Quantum error correction performance against errors on all qubits, averaged over 6 input states}.
This figure is a complement to figure 5 in the main text. It presents results from the same experiment but differs in that the average teleportation fidelity is now averaged over 6 input states (compared to 4 previously). The two new input states are eigenstates of the error process ($Z$ errors). See text for more details. 
\textbf{a} - \textbf{c}. Clusters $\ket{EC_n}$ for  $n=$ 1, 3 and 5, respectively. Rotations (errors) $R_z=\exp(-i\frac{\theta}{2}Z)$ are applied to all central qubits (in blue), which after measurement are projected into independent Z flip errors with probability $p{=}\sin^2(\theta/2)$.
\textbf{d} Ideal performance for cases \textbf{a} - \textbf{c} shown as solid blue, red and black lines, respectively. Increasing the codeword length ($n$) improves performance for $p\leq0.5$.
\textbf{e} Experimental results for cases \textbf{a} - \textbf{c} shown as blue diamonds, red squares and black triangles, respectively.}
\label{all6_global}
\end{figure}

Figures 4 and 5 in the main text present results for our measurement-based quantum error correction demonstration. In both cases the performance is quantified by the average teleportation fidelity (ATF), where the average is taken over the four input states  $\ket{+}{=}(\ket{0}{+}\ket{1})\sqrt{2}$, $\ket{-}{=}(\ket{0}{-}\ket{1})\sqrt{2}$, $\ket{i}{=}(\ket{0}{+}i\ket{1})\sqrt{2}$ and $\ket{-i}{=}(\ket{0}{-}i\ket{1})\sqrt{2}$. Figures \ref{all6_local} and \ref{all6_global} in this text present results, from the same experiments shown in Figures 4 and 5 in the main text, respectively, but now the ATF is taken over six input states. These six are the previous four plus the eigenstates of the Pauli $Z$ operator: $\ket{0}$ and $\ket{1}$. Ideally these two new states should be perfectly teleported across the graph-states, independent of the error probability $p$, since they are eigenstates of $Z$ errors. Experimentally we find that these states have an imperfect fidelity due to the imperfect preparation of the graph-states. These additional states do not probe the error correction capability of the graphs.

The conclusion that we draw from Figure \ref{all6_local} is the same as in the main text: the graphs respond correctly to errors applied to specific codeword qubits ($C_n$) and the larger graphs are able to cope with errors on multiple qubits. The conclusion that we draw from Figure \ref{all6_global} is also unchanged: there is a region where the larger codewords perform better, in spite of the increased complexity of their generation. The size of this region is smaller when including all 6 states. 

Figures 4 and 5 in the the main text show deviations from the ideal cases. We looked into this is in detail and now report on two observations, which are described below. 

\begin{figure}[tb]
\centering
\includegraphics[width=1\columnwidth]{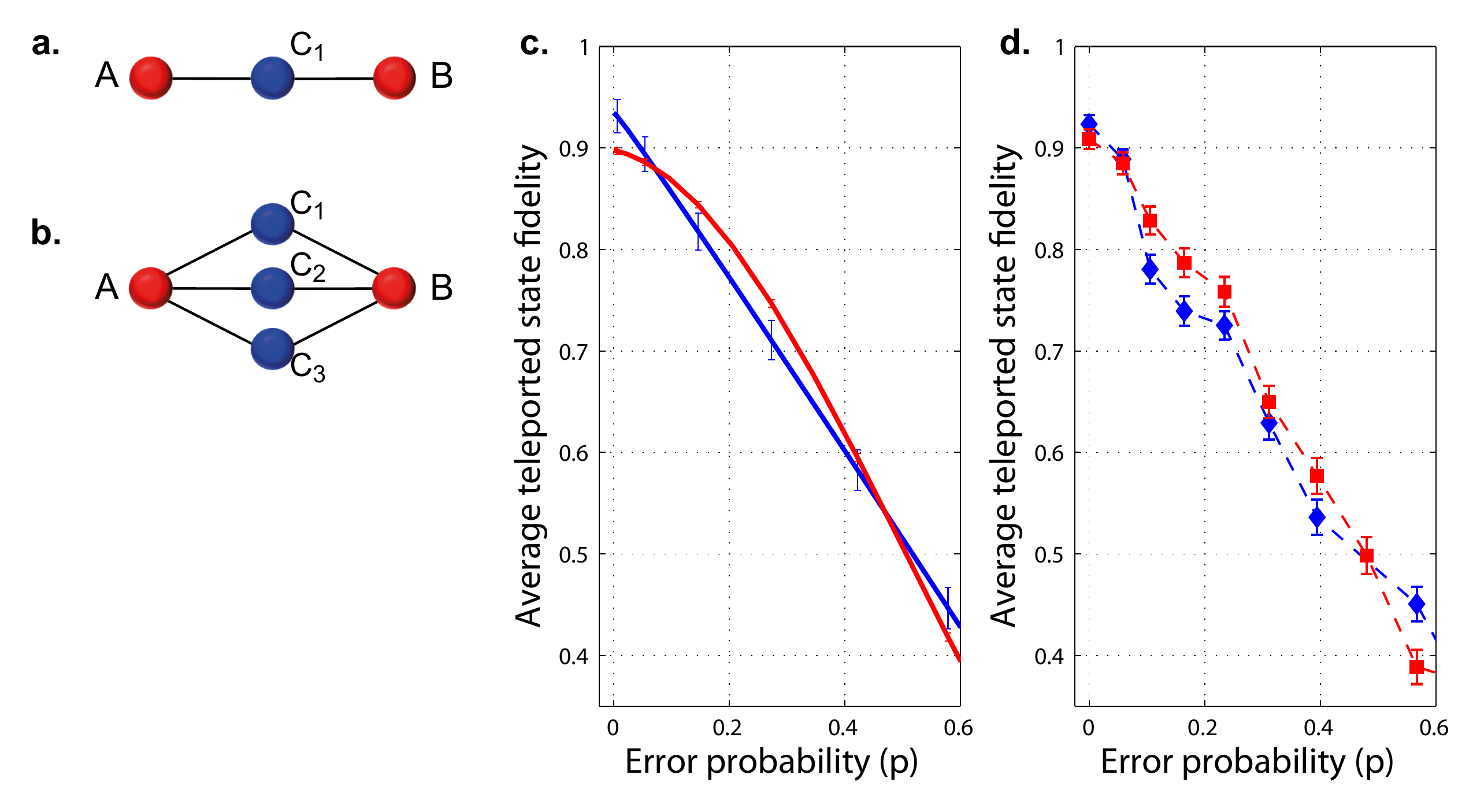}
\caption{\textbf{Quantum error correction performance against errors on all qubits, derived from tomographically reconstructed density matrices}.
This figure is a complement to figure 5 in the main text. Panels a), b) and d) present the same results as in figure 5 in the main text, for the 3 and 5 qubits error correction graphs when errors are applied to all codeword qubits. Panel c) shows results from the same protocol, but derived from experimentally reconstructed density matrices for these graph states. 
\textbf{a} - \textbf{b}. Clusters $\ket{EC_n}$ for  $n=$ 1 and 3, respectively. Rotations (errors) $R_z=\exp(-i\frac{\theta}{2}Z)$ are applied to all central qubits (in blue), which after measurement are projected into independent Z flip errors with probability $p{=}\sin^2(\theta/2)$.
\textbf{c} Performance calculated from experimentally reconstructed density matrices of the state in a) solid blue line and  b) solid red line. 
\textbf{d} Experimental results by directly measuring the states in a) blue diamonds and b) red squares directly.}
\label{err_rho}
\end{figure}

Firstly, since full density matrices of the $n=1$ and $n=3$ graph states were measured, these can be used to simulate the outcome of the error correction tests that we performed. Specifically the entire error correction protocol, including the addition of errors, can be straightforwardly carried out by appropriately rotating and projecting the reconstructed density matrices. The results calculated in this way are presented in figure \ref{err_rho} c and compared to the results that are presented in the main text. The different approaches show a close agreement. From this we can conclude that the main sources of deviation between experiment and ideal theory are due to the errors in the experimentally generated graph states themselves, rather than mistakes in the way we precisely apply errors. 

The high fidelity of the reconstructed states reflects the small errors they contain. Error budgets for two ionic-qubit logic gates have been given before \cite{Benhelm:2008uq}. Specific and detailed experiments will have to be design to determine the extent of potential noise sources in multi-qubit operations. However, it is clear that contributing error sources are laser intensity fluctuations and random electric fields that lead to am incoherent heating rate of the ion string common vibrational modes. We observe a heating rate of approximately 50 phonons per second, of the axial COM at 1.2~MHz.

Secondly, it was surprising to us that the average teleportation fidelity for the $n=5$ state, at $p=0$, is not significantly lower than for the smaller states: this larger 7-qubit state is certainly experimentally more challenging to generate.  We looked into the effect of adding various types of noise to the ideal graph states on the teleportation fidelity at $p=0$ (i.e. without additional $Z$ errors). We found that the larger states are more robust to various types of noise, such as depolarisation. This may be understood by considering that any component of noise that leads to $Z$ errors is better protected by the larger states. Indeed, one should expect that any part of the experimental noise, introduced in our graph-state generation process, that is equivalent to $Z$ errors on the ideal state is better corrected by the error correction protocol for the larger states.

\section{Characterisation of experimentally generated graph states}

\subsection{Quantum state tomography}

We perform full quantum state tomography of all experimentally generated graph states, except the 7-qubit state $\ket{EC_5}$ as the number of measurements required for complete state tomography in this case is impractical. Nevertheless we still confirm the non-classical nature of the 7-qubit state via a multipartite Bell inequality, described in the next section.

Density matrices are reconstructed, from experimentally estimated expectation values, using the maximum likelihood method 
\cite{PhysRevA.64.052312}. We measure an over-complete set of expectation values corresponding to all $3^n$ combinations of tensor products of the Pauli operators. The experimentally reconstructed density matrices, which are ideally states $\ket{E_{LC4}}$, $\ket{E_{EC1}}$, $\ket{E_{EC3}}$ and $\ket{E_{RC4}}$, are presented in figures \ref{LC4}, \ref{EC1}, \ref{EC3} and \ref{RC4}, respectively. From these density matrices, properties such as the ideal-state fidelity and purity are calculated (see table 1 in the main text).  Errors are determined via Monte Carlo (MC) simulation of projection noise (due to a finite number of measurements) centred around the experimentally estimated probabilities \cite{PhysRevA.64.052312}. The MC simulations produce a distribution of `noisy' density matrices, from which uncertainties in derived properties can be estimated.

\section{Multipartite Bell inequalities}

Table 1 in the main text presents results from measurements of multipartite bell inequalities. We were able to show that all our experimentally generated graph states violate such an inequality. We now give more details on this. 

Let $K_i=X_i \prod_{j \in N(i)}Z_j$ denote the stabilizing operators of an $n$-qubit graph state $\ket{G}$ associated with the graph $G$. The group of their products, $S(G)$, is called the stabilizer \cite{PhysRevA.54.1862}. It is given by

\begin{equation}
S(G)= \{s_j , j=1,..., 2^n \}
\end{equation}
with $s_j=\prod_{i \in I_j(G)}K_i$ where $I_j(G)$ denotes a subset of the vertices of $G$. The normalized Bell operator is then defined as 

\begin{equation}
{\cal{B}}_n(G)=\frac{1}{2^n} \sum_{i=1}^{2^n}s_i(G).
\end{equation}

Let ${\cal{D}}(G)=\operatorname{max}_{\operatorname{LHV}}|\langle{\cal{B}}_n\rangle|$ where the maximum is taken over all local hidden variable (LHV) models. Any graph state $\ket{G}$ fulfills $\bra{G}{\cal{B}}_n(G)\ket{G}=1$. Consequently one obtains a Bell inequality whenever ${\cal{D}}(G) < 1$. It reads $\bra{G}{\cal{B}}_n(G)\ket{G} > {\cal{D}}(G)$.

As an example, the normalized Bell operator for the graph $LC_4$ is given by

\begin{eqnarray}
{\cal{B}}_4(LC_4) & = & \frac{1}{16} (\mathbb{I}\mathbb{I}\mathbb{I}\mathbb{I} + XZ\mathbb{I}\mathbb{I} + ZXZ\mathbb{I} + \mathbb{I}ZXZ \nonumber \\
 &  & {} + \mathbb{I}\mathbb{I}ZX + YYZ\mathbb{I} + X\mathbb{I}XZ + XZZX \nonumber \\
 &  & {} + ZYYZ + ZX\mathbb{I}X + \mathbb{I}ZYY - ZYXY \nonumber \\
 &  & {} + X\mathbb{I}YY + YY\mathbb{I}X - YXYZ + YXXY).
\end{eqnarray}

The value ${\cal{D}}(LC_4)$ is given by $0.75$ \cite{PhysRevLett.95.120405}. In the experiment we find $\langle {\cal{B}}(LC_4) \rangle_{exp} = 0.85 \pm{0.02}$ and thus a violation of the Bell inequality by many standard deviations (see table 1 in the main text).

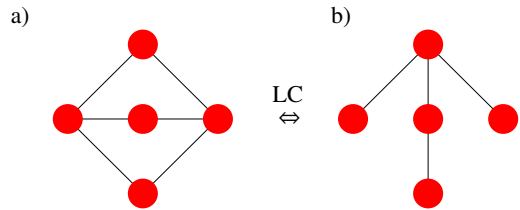
\begin{figure}[htb]
\centering
\includegraphics[scale=0.35]{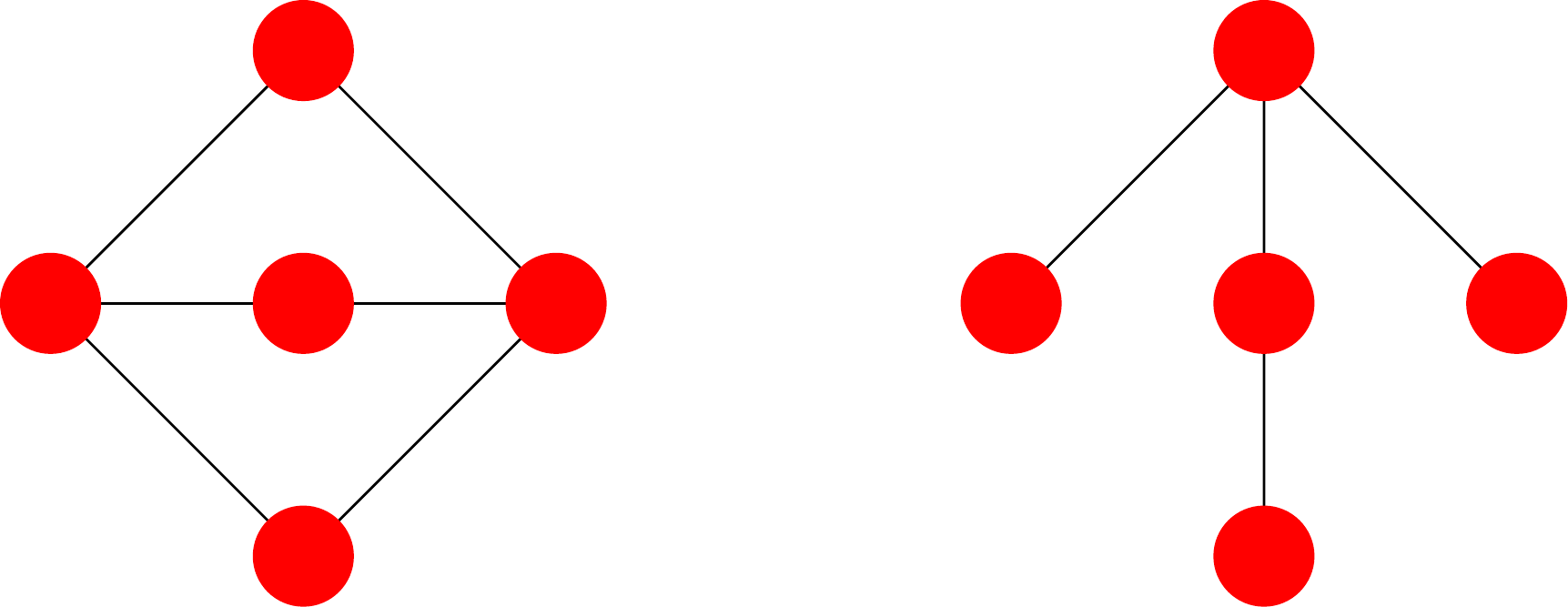}
\put(-190,70){a)}
\put(-70,70){b)}
\put(-90,32){$\Leftrightarrow$}
\put(-92,40){LC}
\caption{The two graph states are LC equivalent, that is they differ only by local Clifford operations. a) $\ket{EC_3}$ b) $\ket{EC_{3LC}}$.}
\label{bell5}
\end{figure}

\begin{figure*}[htb]
\centering
\includegraphics[width=1.4 \columnwidth]{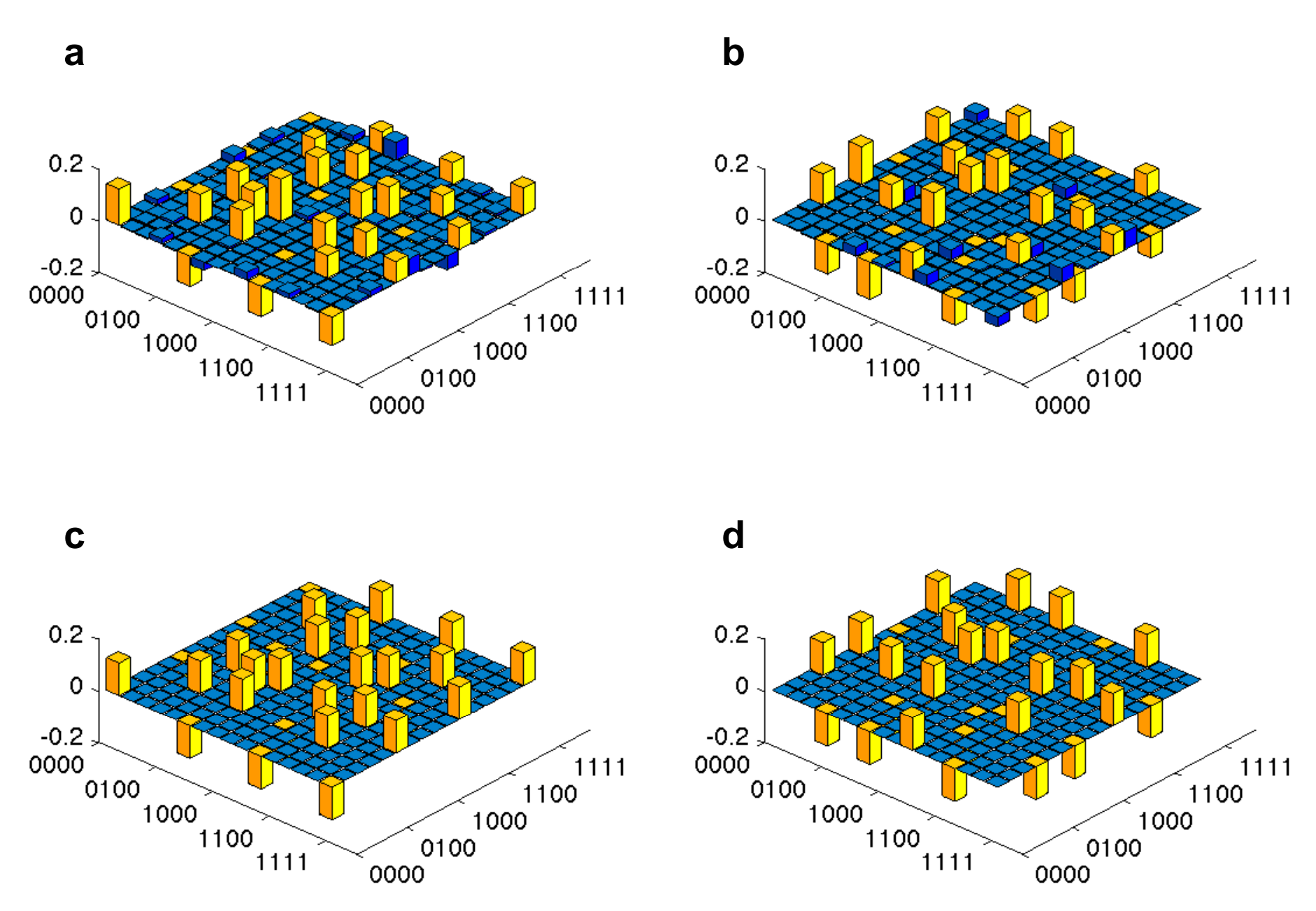}
\caption{\textbf{Density matrix of the 4-qubit linear cluster $\ket{E_{LC_4}}$}.  a) Real and b) imaginary parts of the experimental density matrix, reconstructed via maximum likelihood estimation \cite{PhysRevA.64.052312}. c) Real and d) imaginary parts of the ideal density matrix. Elements with an absolute value greater than 0.1 in the ideal state are shown in yellow. The overlap (fidelity) between the experimental and ideal case is 0.841$\pm{0.006}$. 
}
\label{LC4}
\end{figure*}

\begin{figure*}[htb]
\centering
\includegraphics[width=1.4 \columnwidth]{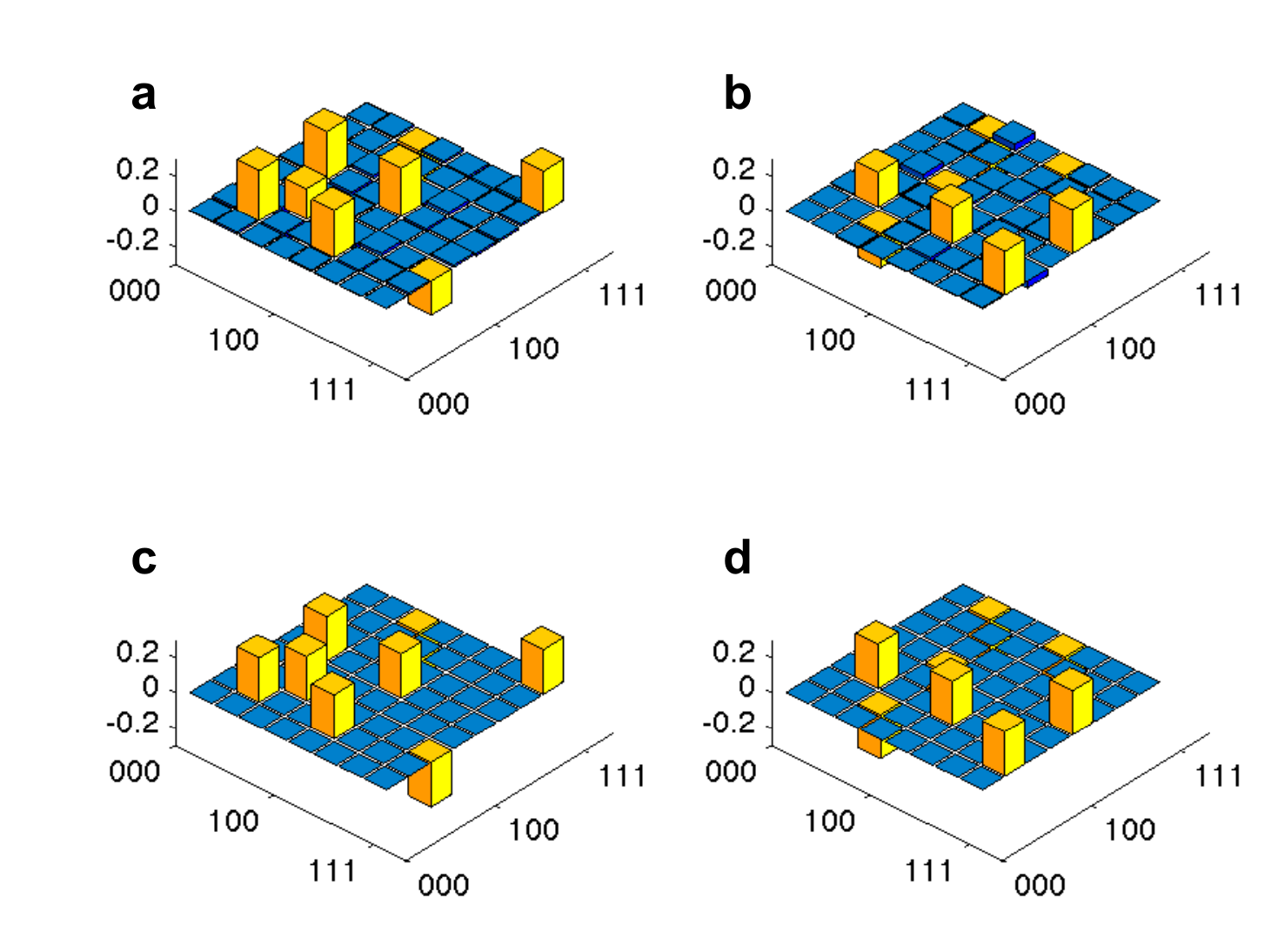}
\caption{\textbf{Density matrix of the 3-qubit linear cluster $\ket{E_{EC_1}}$}.  a) Real and b) imaginary parts of the experimental density matrix, reconstructed via Maximum Likelihood estimation \cite{PhysRevA.64.052312}. c) Real and d) imaginary parts of the ideal density matrix. Elements with an absolute value greater than 0.1 in the ideal state are shown in yellow. The overlap (fidelity) between the experimental and ideal case is 0.920$\pm{0.005}$. 
}
\label{EC1}
\end{figure*}

\begin{figure*}[htb]
\centering
\includegraphics[width=1.7 \columnwidth]{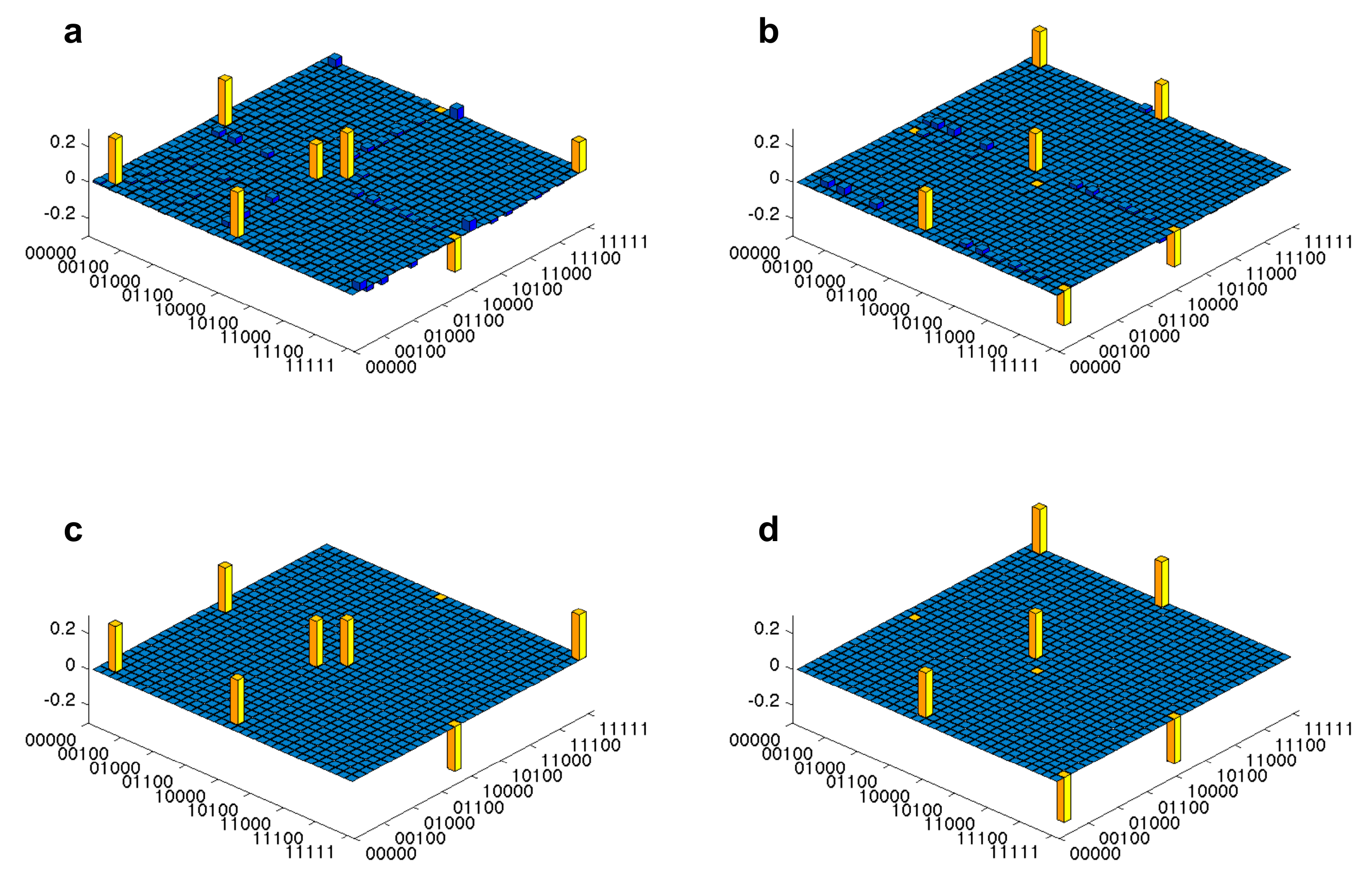}
\caption{\textbf{Density matrix of the 5-qubit linear cluster $\ket{E_{EC_3}}$}.  a) Real and b) imaginary parts of the experimental density matrix, reconstructed via Maximum Likelihood estimation \cite{PhysRevA.64.052312}. c) Real and d) imaginary parts of the ideal density matrix. Elements with an absolute value greater than 0.1 in the ideal state are shown in yellow. The overlap (fidelity) between the experimental and ideal case is 0.843$\pm{0.005}$. 
}
\label{EC3}
\end{figure*}

\begin{figure*}[htb]
\centering
\includegraphics[width=1.7 \columnwidth]{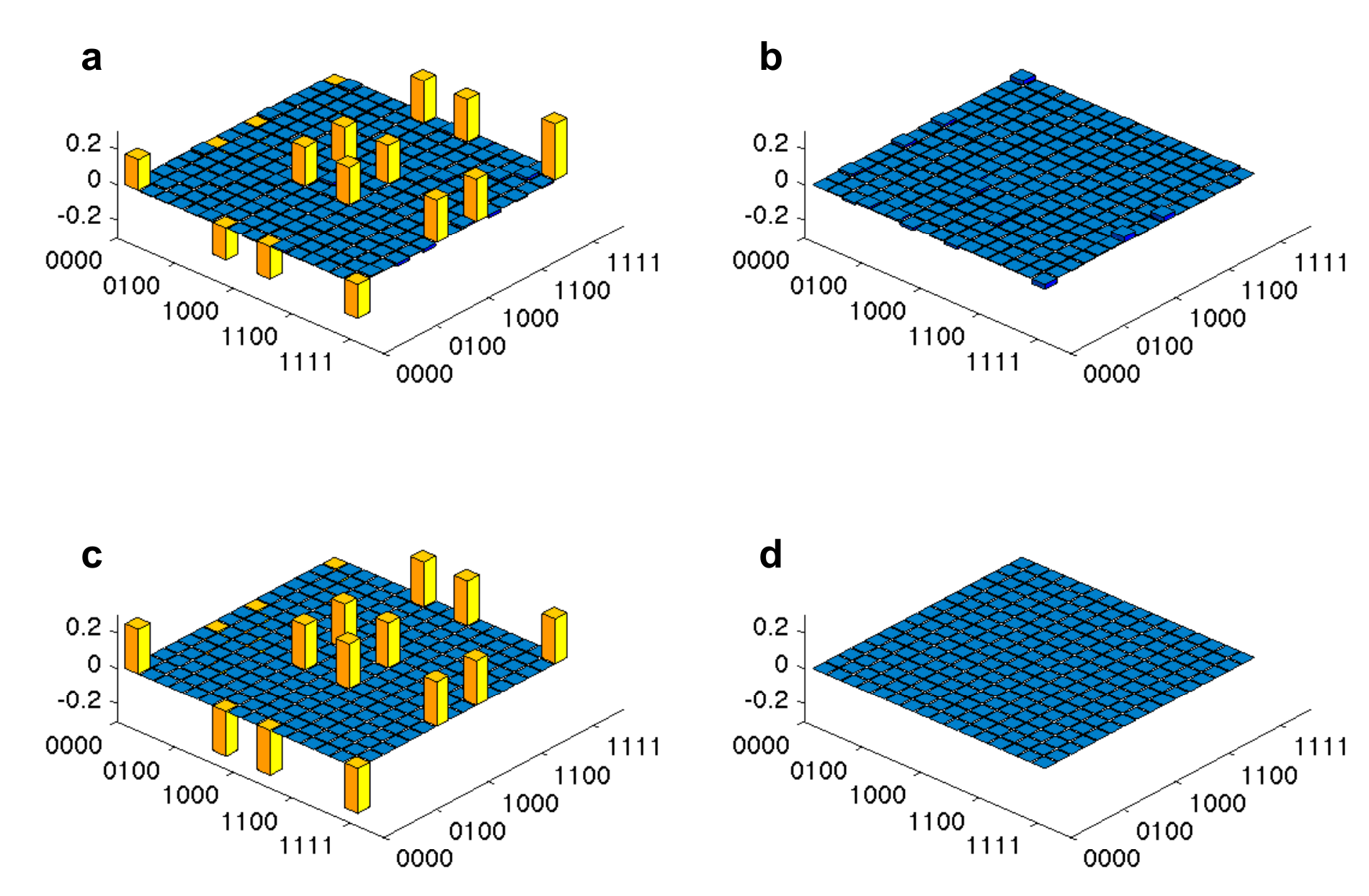}
\caption{\textbf{Density matrix of the 4-qubit ring cluster $\ket{E_{RC_4}}$}.  a) Real and b) imaginary parts of the experimental density matrix, reconstructed via Maximum Likelihood estimation \cite{PhysRevA.64.052312}. c) Real and d) imaginary parts of the ideal density matrix. Elements with an absolute value greater than 0.1 in the ideal state are shown in yellow. The overlap (fidelity) between the experimental and ideal case is 0.847$\pm{0.007}$. 
}
\label{RC4}
\end{figure*}

The value of ${\cal{D}}(G)$ for the three and four qubit linear cluster as well as for the box cluster state can be found directly in \cite{PhysRevLett.95.120405}. A bound for ${\cal{D}}(EC_3)$, where $EC_3$ is the graph underlying the five qubit state $\ket{EC_3}$ (used to demonstrate quantum error correction), can be found in the following way. First one notes that $\ket{EC_3}$ is equivalent to the state $\ket{EC_{3LC}}$ in figure \ref{bell5} b) up to local Clifford (LC) operations. The two graph states have the same rank indices and are thus equivalent up to local unitary operations \cite{PhysRevA.69.062311}. The fact that they are both graph states then implies the LC equivalence. The local Clifford operations do not change the value of ${\cal{D}}(EC_3)$. The graph state $\ket{EC_{3LC}}$ is built from a four qubit GHZ state $\ket{GHZ_4}$ and a single qubit graph $\ket{G_1}$, connected by an edge. Application of Lemma 3 in \cite{PhysRevLett.95.120405} then gives a bound on ${\cal{D}}(EC_3)$:

\begin{equation}
{\cal{D}}(EC_3) \leq {\cal{D}}(G_1) {\cal{D}}(GHZ_4) = \frac{3}{4}.
\end{equation}

In a similar way one can bound the value ${\cal{D}}(EC_5)$,

\begin{equation}
{\cal{D}}(EC_5) \leq {\cal{D}}(G_1) {\cal{D}}(GHZ_6) = \frac{5}{8}.
\end{equation}

A straightforward calculation shows that the normalized Bell operator equals the projector onto the graph state:

\begin{equation}
{\cal{B}}_n(G) = \frac{1}{2^n} \sum_{i=1}^{2^n} s_i = \ket{G}\bra{G}.
\end{equation}

Consequently, one can determine the fidelity $F(\rho_{G_{exp}}) = \operatorname{Tr}(\rho_{G_{exp}}\ket{G}\bra{G})$ of an experimentally obtained graph state with density matrix $\rho_{G_{exp}}$ by measuring the normalized Bell operator. 

For all graph states, except the 7-qubit state $\ket{E_{EC5}}$, we have the state fidelity measured in two different ways: 1. From the reconstructed density matrix and 2. from $\langle {\cal{B}}_n\rangle$. The results are presented in the table in the main text. Both ways give the same results to within statistical uncertainty of one or two standard deviations. One should not expect the values to be exactly the same since only a subset of the tomographic measurements are used in calculating $\langle {\cal{B}}_n\rangle$, giving different statistical contributions.

\section{Ring cluster state}

We created the 4-qubit box cluster state $\ket{RC_4}$, which is an example of a ring cluster \cite{PhysRevLett.95.120405} and the smallest intrinsically 2D cluster state. The state $\ket{RC_4}$ can be generated by preparing all qubits in $\ket{+}$ and applying CPHASE gates between every pair of qubits connected by an edge. i.e. $\ket{RC_4}=C_{14}C_{34}C_{23}C_{12}\ket{+}^{\otimes n}$, where $C_{ij}$ is a CPHASE between qubits $i$ and $j$.

We experimentally generate a state $\ket{E_{RC4}}$ which is equivalent to $\ket{RC_4}$ up to single-qubit unitary operations (i.e. they are locally equivalent). Specifically, the local unitary operators, which have to be applied to $\ket{E_{RC4}}$ to convert it to $\ket{RC_4}$ are shown in table \ref{tablecorbox}. The laser pulse sequence used to generate the state $\ket{E_{RC4}}$ is the same as that to generate the error correction graph states, and is shown in figure \ref{gates_EC}.

%\bibliography{clusters}

\newpage

\end{document}